\documentstyle[preprint,aps]{revtex}
\begin{document}
\draft

\title{
{\bf ANALYTIC SOLUTIONS OF QCD EVOLUTION EQUATIONS FOR
PARTON CASCADES INSIDE NUCLEAR MATTER AT SMALL X}
}

\author{
{\bf K. Geiger} \\
{\it CERN TH-Division, CH-1211 Geneva 23, Switzerland}
}

%\date{\today}

%\maketitle
\setcounter{page}{0}

\begin{flushright}
CERN-TH. 7206/94
\end{flushright}

\begin{center}
{\Large
{\bf ANALYTIC SOLUTIONS OF QCD EVOLUTION EQUATIONS FOR
PARTON CASCADES INSIDE NUCLEAR MATTER AT SMALL X}
}
\end{center}
\bigskip

\begin{center}
{\bf K. Geiger}

{\it CERN TH-Division, CH-1211 Geneva 23, Switzerland}
\end{center}

\begin{abstract}
An analytical method is presented to solve generalized QCD evolution
equations for the time development of parton cascades in a
nuclear environment.
In addition to the usual parton branching processes in vacuum,
these evolution equations provide a consistent description of
interactions with the nuclear medium by accounting for
stimulated branching processes, fusion and scattering processes
that are specific to QCD in medium.
Closed solutions for the spectra of produced partons with respect to the
variables time, longitudinal momentum and virtuality are
obtained under some idealizing assumptions about the composition
of the nuclear medium. Several characteristic features of the resulting
parton distributions are discussed. One of the main conclusions is that
the evolution of a parton shower in medium is dilated as compared
to free space and is accompanied by an  enhancement of particle
production. These effects become stronger with increasing nuclear
density.
\end{abstract}
\noindent

\vspace{0.5cm}

\leftline{PACS Indices: 25.75.+r, 12.38.Bx, 12.38.Mh, 24.85.+p}
\rightline{e-mail: klaus@surya11.cern.ch}
\leftline{CERN-TH. 7206/94,  March 1994}

\vfill \eject

\noindent {\bf 1. INTRODUCTION}
\medskip

In a preceding paper \cite{gm1}, B. M\"uller and myself
addressed the question,
how is ``QCD in medium'' modified as
compared to ``QCD in vacuum'' when one attempts to extend
well developed perturbative QCD techniques \cite{muellbook}
from high energy hadron-hadron collisions
to ultra-relativistic heavy ion collisions.
The investigation of \cite{gm1} condensed in the derivation of
a generalized form of QCD evolution equations that
describe the time evolution of parton distributions in a nuclear medium,
solutions of which I will present in this paper.
This aspect is of great importance for the future
ultra-relativistic heavy ion collider experiments at
the BNL Relativistic Heavy Ion Collider (RHIC)  and
the CERN Large Hadron Collider (LHC), where
new phenomena associated with ``QCD in medium'' are expected to
clearly modify naive extrapolations from hadron-hadron collisions.
Although in the last couple of years, numerical simulations with
QCD based Monte Carlo models \cite{pcm0,pcmapp,hijing,dtujet}
have provided considerable new insight in the microscopic
parton dynamics of ultra-relativistic heavy ion collisions,
one is still far from a complete picture and a truly quantitative
description. Aside from the rather intransparent complexity of these
computer simulations, one still relies on a substantial amount
of phenomenological modeling of nuclear and dense medium effects
due the current lack of better knowledge about the details of
such mechanisms.
This is reflected by a very large uncertainty in quantitative
predictions: although the different model calculations generally agree very
well for $pp$ collisions at collider energies,
they differ in their predictions,
for e.g. charged particle multiplicities, in heavy ion $AA$ collisions
by a factor of 2 or more.

These issues lead to the main motivation for this work, namely the necessity
to go back to a simpler, but more fundamental and transparent level, and
study within the firmly established framework of perturbative QCD
(rather than extending phenomenological model input)
the microscopic dynamics of quarks and gluons in the hot,
ultra-dense environment that may be created in heavy ion collisions
\cite{qm,qgp,lbl93}.
In order to do so, it is inevitable to disentangle
the various nuclear and medium effects from each other by
investigating physical situations where such medium phenomena may be
singled out, both in theoretical and experimental research.
For instance, deep inelastic scattering on heavy
nuclei \cite{frankstrik},
charmonium production by Drell-Yan processes in nuclei \cite{karsatz},
energy loss of partons, jet quenching  and multiple scattering
phenomena in hadron-nucleus collisions
\cite{GW91}, etc., provide opportunities to explore nuclear
modifications to the QCD parton picture that has been so successful
in high energy particle physics.

As a first step in this direction, in Ref. \cite{gm1}
a generalized form of the well known
Dokshitzer-Gribov-Lipatov-Altarelli-Parisi (DGLAP) evolution
equations \cite{dok77,ap,dokbook} was derived. In the framework of the
leading logarithmic approximation (LLA) \cite{lla1,lla2,lla3,lla4},
we obtained a set of coupled integro-differential
equations for the {\it time evolution} (rather than the $Q^2$-evolution)
of the off-shell parton distribution functions in nuclear matter.
In addition to the usual parton branching processes in vacuum,
these evolution equations provide a self-contained description of
interactions with the nuclear medium by accounting for
stimulated branching processes, fusion and scattering processes,
which are specific to QCD in medium.
This approach resulted in a probabilistic description within perturbative QCD
of the time variation of parton distributions as in
non-equilibrium kinetic theory, however now
including consistent treatment of off-shell propagation of partons,
a concept that is absent in semiclassical transport theory \cite{degroot}.
The essential condition for the applicability of perturbative
techniques is that the QCD factorization theorem \cite{colsop}
holds also in the presence of dense nuclear matter, in which multiple
interactions of the evolving partons with this nuclear medium
may be separated in space-time by only very short distances.
Recent investigations by
McLerran and Venugopalan \cite{MLV93} are encouraging in this direction,
since they showed that
a consistent perturbative calculation of parton structure functions
at small values of the Bjorken variable $x$ becomes possible when one
considers the limit of a very thick nuclear target.
The condition for the applicability of perturbative QCD is then that
medium induced effects, such as color screening and rescattering,
provide dynamical cut-offs on a scale short compared to the
QCD renormalization scale $\Lambda^{-1}$.

Exploring this insight, the present paper as a sequel to the work of
Ref. \cite{gm1} employs  the probabilistic
parton cascade  picture \cite{dokbook,qm93}
to describe  the evolution of  a parton shower inside nuclear matter,
triggered by a highly excited initial quark or gluon.
Here I will start from the generalized evolution equations and
solve for the parton spectra with respect to the
variables time, longitudinal momentum and virtuality,
under some idealizing assumptions about the composition
of the nuclear medium.
Before going to the heart of the matter, let me remark the following:
The approach is at this point in a yet idealized stage,
since various unsettled problems are left out, and a
number of simplifying approximations are made.  Still, it is
a starting point which can be step by step improved, similarly maybe
as the development of perturbative QCD in hard processes that
started some 15 years ago.
The main assumptions and approximations in Ref. \cite{gm1} and in
the present paper can be stated as follows:
\begin{description}
\item[(i)]
Factorization of short distance interactions of shower partons
with the nuclear medium from long range non-perturbative forces is
assumed.
This means, even in the presence of dense medium where a parton
can encounter multiple successive interactions, a probabilistic
description of local, non-interfering interactions applies at
sufficiently high energies.
\item[(ii)]
Neglect of interference effects that affect particularly the
softest partons: destructive interference of interaction
amplitudes in both, coherent succesive small angle
emissions ('angular ordering'), and multiple sequential scatterings
('Landau-Pomerantchuk-Migdal effect').
\item[(iii)]
Nuclear effects due to long range correlations, e.g. nuclear
shadowing (antishadowing) not associated with truly perturbative
parton interactions are ignored.
\end{description}

The remainder of the paper is organized as follows.
Sec. 2 is devoted to recall the intuitive picture developed
in Ref. \cite{gm1} of the probabilistic
parton evolution in medium. The theoretical framework is reviewed and the
master equations describing the time development of the distribution of
off-shell partons in a nuclear environment are summarized.
Sec. 3 then deals with linearizing and solving these equations analytically by
making a number of simplifying assumptions regarding the nuclear medium.
Explicit closed solutions for parton showers initiated by an energetic
time-like quark or gluon are obtained, which in the special case
of absence of nuclear matter reduces to the well known
LLA solutions for parton distributions, however now
time being the evolution variable rather than virtuality.
Some general features concerning the $x$-dependence as well as the
influence of the nuclear density are discussed.
A summary and outlook is given in Sec. 4.
\bigskip

\noindent {\bf 2. THE COUPLED EVOLUTION EQUATIONS FOR QUARKS,
ANTIQUARKS  AND GLUONS IN MEDIUM}
\medskip

Let me start by recalling the essential considerations of
Ref. \cite{gm1} that lead to the derivation of a
coupled set of integro-differential equations for a
system of off-shell quarks (antiquarks) $q_i$
($\bar q_i$) with flavors $i=1, \ldots , n_f$, and gluons $g$.
As will become clear, these Boltzmann type equations
describe the time evolution of parton cascades in
a nuclear medium with account of off-shell
propagation of time-like virtual partons.
Although as mentioned before, the evolution equations may be
applied to a variety of physical situations, let me specify
here for the purpose of lucidity a definite, though somewhat
idealized scenario:
Consider a fast parton injected into infinitely extended nuclear
matter by some highly localized process of space-time extent,
$(Q_0^2)^{-1/2} \ll \Lambda^{-1}$.
This {\it primary} parton then propagates through the
nucleus and initiates a cascade or
shower of {\it secondary} partons. The partons in the cascade
can either radiate bremsstrahlung gluons or produce $q\bar q$-pairs,
collide with partons of the nucleus (the medium),
or absorb nuclear partons.  I will call such an event
a {\it parton shower} or {\it parton cascade}.  I emphasize that I
will explicitly distinguish between the {\it shower partons} on the one
hand, and the {\it nuclear partons} on the other hand which initially are
coherently bound in the wave function of the nucleus.
Also, for the time being, I consider here
the evolution of a single
cascade and therefore only account for interactions of the cascade with
the medium, but neglect possible interactions between simultaneously evolving
cascades.

The prototype reaction of such a scenario is a proton-nucleus ($pA$) collision
at very high beam energy, where one may trigger on a high energy time-like
quark or gluon that is produced at some point of time $t_0$
inside the heavy nucleus by a hard scattering, and single out the
development of the resulting parton shower inside the nucleus
with a sufficiently homogenous spatial density.
For the description of such a parton cascade
it is convenient to choose the {\it nucleon-nucleon center-of-mass frame}
(CM$_{NN}$)  \cite{pcmpa} in which each nucleon has the
same value of longitudinal momentum $P$,
\begin{equation}
P_z^{(p)} \;=\; + \, P\;\;,\;\;\;\;\;\;\;
P_z^{(A)} \;=\; - \, A\, P
\;\;,\;\;\;\;\;\;\;
\vec P_\perp^{(p)} = \vec P_\perp^{(A)} = \vec 0
\;\;,
\label{e1}
\end{equation}
with
\begin{equation}
\sqrt{s}\;=\; \sqrt{4 A P^2 \;+\; M_N^2(1 + A^2)}
\;\;, \;\;\;\;\;\;\;\;\;\;\;
\sqrt{s_{NN}}\;\equiv\; 2\;P
\;\;,
\end{equation}
where $A$ is the nuclear mass number, $M_N$ the nucleon mass, and
$P/M_N \gg 1$ is assumed. For example, at the RHIC (LHC) the
maximum available beam energy implies $P/M_N >$ 100 (3000) even for $p+Au$
collisions. I will describe the longitudinal evolution of a parton shower along
the {\it shower axis} ($z$-axis), which I define parallel to the
direction of momentum of the initiating primary parton.  It is
convenient to parametrize the four-momenta $k\equiv k^\mu = (E, k_z,
\vec k_\perp)$ of the shower partons such that for the primary parton,
\begin{equation}
k_0\;=\;\left(x_0 P+\frac{Q_0^2}{2 x_0 P}; \;x_0 P, \;\vec 0 \,\right)
\;\;,
\label{p0}
\end{equation}
whereas for the $j^{th}$ secondary parton,
\begin{equation}
k_j\;=\;\left(x_j P+\frac{Q_j^2+k_{\perp j}^2}{2 x_j P}; \;x_j P, \;
\vec k_{\perp j} \right) \;\;,
\label{pj}
\end{equation}
where $x_j= p_{z\,j}/P$
can be either positive or negative depending on the
partons direction of propagation along the $z$-axis.
Furthermore, $k_{\perp j}^2 < Q_j^2  \ll P^2$ is assumed, and all
rest masses are neglected.  It is important to realize that energy and
momentum are independent variables, since one is dealing with off-shell
particles of virtuality $Q^2>0$ with a continuous invariant mass
distribution.
The evolution of the parton system is then described by the change of
{\it parton number densities}, which are defined as
\begin{equation}
a(x, k_\perp^2; Q^2, t)\;=\; \int_{0}^{Q^2}  d Q^{\prime \, 2}\,
\frac{d N_a(t)}{d x d k_\perp^2 d Q^{\prime\,2}}
\;\;,\;\;\;\;\;\;(\,a\,\equiv \,q_i, \bar q_i, g\,;
\;\;i = 1, \ldots , n_f )
\;\;,
\label{axt}
\end{equation}
or, when integrated over transverse momentum,
\begin{equation}
a(x, Q^2, t)\;=\; \int d k_\perp^2
\; a(x, k_\perp^2; Q^2, t)
\;\;.
\label{axt1}
\end{equation}
The relation between the parton number densities (\ref{axt}) to the
single-particle phase-space distributions is expressed by
the number of partons of type $a$ present at time $t$
\begin{equation}
N_a(t)\;=\;\int d^3 r \int d^4 k\,F_a(E,\vec k; t, \vec r)
\;\equiv\; \int d^4 k\,f_a(E,\vec k; t)
\;\;,
\end{equation}
with $F_a$ denoting the corresponding phase-space density of
off-shell partons with virtuality $Q^2=E^2 - \vec k ^2$
in the phase-space volume $dE d^3k d^3r$ around $k^\mu$ and
$\vec r$ at time $t$, and $f_a$ in the second line
representing the spatially integrated energy-momentum distribution.

In order to describe the time evolution of the parton densities
(\ref{axt}) or (\ref{axt1}) one has to relate the
change of the dynamical variables $x$ and $Q^2$ with the
laboratory time $t$ which plays the role of an external 'parameter'
rather than being an intrinsic kinematical quantity.
In Ref. \cite{gm1} it was shown, using time-dependent perturbation
theory, that for the case of a parton shower evolving in
vacuum (i.e. by successive branchings only), one finds
that the time scale for a branching chain, starting at
$t_0=0$ with initial values $Q_0^2$ and $x_0$, is in the average given by
\begin{equation}
t\;=\;\sum_{j=1}^n \,\frac{|x_j|P}{Q_j^2} \;\simeq \; \frac{|x|P}{Q^2}
\label{txp}
\end{equation}
with ($z_j = x_j/x_{j-1}$)
\begin{equation}
x \;=\;x_n \;=\;z_1 z_2 \;\ldots \;z_n x_0
\;\;,\;\;\;\;\;\;\;\;\;\;\;
Q^2\;=\;Q_n^2 \;\ll\;Q_{n-1}^2 \;\ll\; \ldots \;\ll\;Q_0^2
\;\;,
\label{tQ}
\end{equation}
in accord with the uncertainty principle. Thus, in the absence of interactions
with surrounding matter, the time variable $t$ is related to
$x$ and and the degree of off-shellness $Q^2$ by
the typical life-time $1/Q$ in the parton's restframe, boosted by the
Lorentz factor $\gamma = |x|P/Q$.
However, when considering a parton shower inside nuclear matter,
scattering and fusion processes with nuclear partons will compete
with spontaneous branchings and consequently disrupt a steady
decrease of average virtuality with time. Both scatterings and
fusions can increase the virtuality of a parton by energy-momentum transfer.
This means that the relation (\ref{tQ}) between $x$, $Q^2$ and $t$
cannot be deterministic anymore, because interactions with the
nuclear medium occur stochastically according to the density of
nuclear partons.
Each scattering or fusion may "rejuvenate" a shower parton by providing it with
a virtuality $Q^{\prime \,2} > Q^2$ so that the parton after the
interaction appears to be younger ($t < t^\prime$).
To keep track of these repeated rejuvenations, one has to reset the clock
for the particular shower parton after each such interaction with the medium.
For this purpose a new independent dynamic variable $\tau$,
called the {\it age of a parton}, was introduced in Ref. \cite{gm1}.
Instead of (\ref{tQ}), one finds
\begin{equation}
\tau\;\equiv\;\tau(x,t)\;=\;\left(\frac{|x|P}{Q^2}\right)_t
\;\;,
\label{tauQ}
\end{equation}
which determines the typical age of a parton depending on
$x$, $Q^2$ and implicitely also on $t$, and
which introduces an additional time scale that reflects the external
influence of the medium on the time evolution of the parton cascade.
Correspondingly, the parton
number densities (\ref{axt}) and (\ref{axt1})
must be generalized as
\begin{equation}
a(x, \tau, t)\;=\; \int_{\tau_0}^{\tau}  d \tau^{\prime}\,
\int d k_\perp^2 \,\frac{d N_a(t)}{d x  d k_\perp^2 d\tau^{\prime}}
\;\;,
\label{axt2}
\end{equation}
which now has the meaning of the probability density for finding
a parton of type $a$ at time $t$ with age $\tau = |x| P/Q^2$
and momentum fraction $x$. The age variable measures here
the influence of the nuclear medium on the development of the
parton cascade. Note that in absence of a background medium
the age variable looses its independent character, i.e. it then
evolves parallel to laboratory time and $\tau \propto t$.
I will return to that point in Sec. 3.
\medskip

{\bf 2.1 The evolution equations}

\noindent
For the remainder of this paper I will be concerned only with
transverse momentum integrated quantities, so that it suffices to
describe the evolution in terms of
the parton number densities $q_i$, ($\bar q_i$) and $g$,
defined by eq. (\ref{axt2}). In addition, it is convenient to introduce
the {\it parton momentum densities} $Q_i$ ($\bar Q_i$) and $G$
\begin{equation}
Q_i(x, \tau,t) \;\equiv\;
x\,q_i(x, \tau,t) \;\;\;,\;\;\;\;\;\;\;\;\;\;\;
G(x, \tau, t) \;\equiv\;
x \,g(x, \tau, t)
\;\;.
\label{QG}
\end{equation}
i.e. the parton number densities weighted with
the longitudinal momentum fraction $x$, and integrated over all transverse
momenta $k_\perp^2$.

As graphically illustrated in Fig. 1, the
coupled evolution equations for the parton densities $Q_i$ of
quarks, $\bar Q_i$ of antiquarks, and $G$ of gluons, respectively,
can be summarized now as follows \cite{gm1}:
\begin{eqnarray}
\left(\frac{\partial}{\partial t} + \frac{\partial}{\partial \tau}\right)
\, Q_i(x, \tau,t) &=&
- \;\left[
\frac{}{}
\;\hat A  \; + \;
\hat A^\prime\{\tilde G\} \;+\;
\hat B^\prime\{\tilde{\bar{Q}}_i\}  \right] \; Q_i
\,\;+\;\,\left[  \; \frac{}{}\hat B \;+\;
\hat S_{q g}\{\tilde Q_i\} \right]\; G
\nonumber
\\
& &\;\;+\;
\sum_{j=1}^{n_f}
\hat S_{q q}\{\tilde Q_i\} \,\; Q_j \;+\;
\sum_{j=1}^{n_f}
\hat S_{q \bar q}\{\tilde Q_i\} \,\; \bar Q_j
\label{eq}
\\
\left(\frac{\partial}{\partial t} + \frac{\partial}{\partial \tau}\right)
\, \bar Q_i(x, \tau, t) &=&
- \;\left[
\frac{}{}
\;\hat A  \; + \;
\hat A^\prime\{\tilde G\} \;+\;
\hat B^\prime\{\tilde{Q}_i\}  \right] \; \bar Q_i
\,\;+\;\,\left[  \; \frac{}{}\hat B \;+\;
\hat S_{q g}\{\tilde{\bar{Q}}_i\} \right]\; G
\nonumber
\\
& &\;\;+\;
\sum_{j=1}^{n_f}
\hat S_{q \bar q}\{\tilde{\bar{Q}}_i\}\,\; Q_j
\;+\;
\sum_{j=1}^{n_f}
\hat S_{\bar q \bar q}\{\tilde{\bar{Q}}_i\}\,\; \bar Q_j
\label{eqq}
\\
\left(\frac{\partial}{\partial t} + \frac{\partial}{\partial \tau}\right)
\, G(x, \tau, t) &=&
- \;\left[
\;\hat C \;+\;\hat D  \; + \;
\hat C^\prime\{\tilde G\}
\;+\; \sum_{j=1}^{n_f} \left\{ \hat D^\prime\{\tilde Q_j\} \,+\,
\hat D^\prime \{\tilde{\bar{Q}}_j\} \right\}
\;-\;
\hat S_{g g}\{\tilde G\} \right]\; G
\nonumber
\\
& &
\;+\; \sum_{j=1}^{n_f} \left[\frac{}{}
\; \hat E \;+\;
\hat E^\prime \{\tilde{\bar{Q}}_j\}  \;+\;
\hat S_{g q}\{\tilde G\} \right] \; Q_j
\nonumber
\\
& &
\;+\; \sum_{j=1}^{n_f} \left[\frac{}{}
\; \hat E \;+\;
\hat E^\prime \{\tilde Q_j\}  \;+\;
\hat S_{g \bar q}\{\tilde G\} \right]\; \bar Q_j
\;\;.
\label{eg}
\end{eqnarray}
Note that this set of equations describes the time evolution of
the parton (longitudinal) momentum distributions $Q = x q$, $\bar Q = x
\bar q$ and $G = x g$, rather than of the parton number densities $q$,
$\bar q$ and $g$.
The left hand side of these equations describes just the free streaming
in the absence of interactions,
whereas on the right hand side
the various integral operators acting
on the parton momentum densities desribe the change of
the densities due to branching, fusion, and scattering
processes.
In contrast to the branching operators ($\hat A, \hat B,\ldots , \hat E$),
the operators that describe fusions
($\hat A^\prime \{\tilde Q\}, \hat B^\prime \{\tilde Q\},\ldots ,
\hat E^\prime \{\tilde G\}$) and scattering processes
($\hat S_{qq}\{\tilde Q\}, \ldots , \hat S_{gg}\{\tilde G\}$)
are functionals of the nuclear parton distributions
(labeled with a ' $\tilde{} $ ')
of quarks $\tilde Q$ ($\tilde{\bar Q}$) and
gluons $\tilde G$.

In order to represent the integral operators in the evolution
equations (\ref{eq})-(\ref{eg}) in a compact form,
let me introduce effective coupling functions defined in
terms of the running QCD coupling strength in one-loop order
\begin{equation}
\alpha_s (Q^2)\;=\;
\frac{12 \pi}{(33-2 n_f) \ln (Q^2/\Lambda^2)}
\;\;\;,
\label{alpha}
\end{equation}
where $\Lambda$ is the QCD renormalization scale
and $n_f$ is the number of quark
flavors that can be probed at scale $Q^2$. The following definitions
will be employed:
\begin{eqnarray}
\alpha_s\left(\tau\right) &\equiv&
\frac{
\;\int_0^1 d x \;
\alpha_s\left(\frac{\vert x\vert P}{\tau}\right)\;
\theta\left(\frac{\vert x\vert P}{\tau} \,-\,\mu_0^2\right)
\; \left(g\left(x,\tau,t\right)\,+\,\sum_j [q_j(x,\tau,t) +
\bar q_j(x,\tau,t)] \right)
}{
\int_0^1 d x \,\left( g(x,\tau,t) \,+\, \sum_j [q_j(x,\tau,t) + \bar
q_j(x,\tau,t)] \right)
}
\nonumber \\
\nonumber \\
\xi(\tau) &\equiv&
\frac{\alpha_s(\tau)}{2 \pi \tau}
\;\;,\;\;\;\;\;\;\;\;\;\;\;\;\;\;\;
\zeta(x,\tau) \;\equiv\;
\left. {4\pi\alpha_s(M^2)\over M^2} \right|_{M^2 = \vert x\vert P/\tau}
\;\;.
\label{notation}
\end{eqnarray}
Here $\alpha_s(\tau)$ describes the $x$-averaged QCD coupling
at scale $Q^2= \vert x\vert P/\tau$, and the age $\tau$
represents the mean life-time of a parton with longitudinal momentum fraction
$x$ and virtuality $Q^2$.
Finally I denote by $\gamma_{a \rightarrow bc}(z)$
the usual branching functions \cite{dok77,ap},
\begin{eqnarray}
\gamma_{q \rightarrow q g} (z) &=&
C_Q\; \left( \frac{1 + z^2}{1 - z} \right)
\nonumber
\\
\gamma_{q \rightarrow g q} (z) &=&
C_Q\; \left( \frac{1 + (1-z)^2}{z} \right)
\nonumber
\\
\gamma_{g \rightarrow g g} (z) &=&
2\,C_G\;\left( z ( 1 - z ) + \frac{z}{1-z} + \frac{1-z}{z} \right)
\nonumber
\\
\gamma_{g \rightarrow q \bar q} (z) &=&
\frac{1}{2} \, \left( z^2 + (1 - z)^2 \right)
\;\;,
\label{gamma}
\end{eqnarray}
where $C_Q = \frac{n_c^2-1}{2 n_c} = 4/3$, $C_G = n_c =3$,
and $z = x_b/x_a$ is the fraction of $x$-values of daughter to
mother partons in the branching $a \rightarrow bc$.
\medskip

As is evident from the evolutions equations (\ref{eq})-(\ref{eg}),
the rate of time change of the number of partons receives
various contributions from the different interactions of the cascading partons,
represented by their momentum densities $Q$, $\bar Q$ and $G$,
from $1\rightarrow 2$ branchings, $2 \rightarrow 1$ fusions and
$2 \rightarrow 2$ scatterings. In addition to the well known
parton evolution in vacuum (e.g. jet development in $e^+e^-$-annihilation),
in which only spontanous branching processes contribute,
in the present scenario of parton evolution in nuclear matter,
the interactions with the medium,
i.e. with the nuclear partons, give rise to fusion and
scattering processes, and indirectly also to
stimulated branchings. The physical picture underlying this
probabilistic approach to generate such parton cascades from
sequential elementary $1\rightarrow 2$, $2\rightarrow 1$ and $2\rightarrow 2$
interactions is a direct generalization of the standard perturbative QCD
parton shower picture \cite{dokbook,qm93}. For more details I refer to
Ref. \cite{gm1}. Here I will only list the corresponding expressions
of the integral operators in
eqs. (\ref{eq})-(\ref{eg}) corresponding to the diagrams in
Fig. 1.
\smallskip

{\bf 2.2 Branching processes}

\noindent
For cascading quarks
the $1\rightarrow 2$ branching processes are balanced by
the loss ($-$) and gain (+) rates
\begin{eqnarray}
-\;
\hat A \,Q_i &=&
-\;
\int_0^1 d z \left[
Q_i(x,\tau,t) \,-\,Q_i\left(\frac{x}{z},
\tau,t\right)\right]
\;\xi(\tau)
\; \gamma_{q \rightarrow q g}(z)
\nonumber
\\
+\;
\hat B \,G &=&
+\;
\int_0^1 d z
\; G\left(\frac{x}{z}, \tau,t\right)
\;\xi\left(\tau\right)
\; \gamma_{g \rightarrow q \bar q}(z)
\;\;.
\label{bq}
\end{eqnarray}
The corresponding branching rates for antiquarks are obtained upon
substitution $Q_i \leftrightarrow \bar Q_i$.
The loss and gain branching rates of gluons are given by:
\begin{eqnarray}
-\;
\hat C \, G &=&
-\;
\int_0^1 d z \left[
\frac{1}{2}\,
G(x,\tau,t) \,-\,G\left(\frac{x}{z},
\tau,t\right) \right]
\xi(\tau)
\; \gamma_{g \rightarrow g g}(z)
\nonumber
\\
-\;
\hat D \, G &=&
-\;
n_f \;
G(x,\tau,t)
\;
\int_0^1 d z
\;\xi(\tau)
\; \gamma_{g \rightarrow q \bar q}(z)
\nonumber
\\
+\;
\hat E \,Q_j &=&
+\;
\int_0^1 d z
\; Q_j\left(\frac{x}{z},\tau,t\right)
\;\xi (\tau)
\; \gamma_{q \rightarrow q  g}(z)
\nonumber
\\
+\;
\hat E \,\bar Q_j &=&
+\;
\int_0^1 d z
\;\bar Q_j\left(\frac{x}{z},\tau,t\right)
\;\xi (\tau)
\; \gamma_{q \rightarrow q  g}(z)
\;\;.
\label{bg}
\end{eqnarray}
\smallskip

{\bf 2.3 Fusion processes}

\noindent
The $2 \rightarrow 1$ fusion rates that describe the
effective loss of quarks due to fusions are:
\begin{eqnarray}
-\;
\hat A^\prime \{\tilde G\} \,Q_i &=&
-\;
\rho_N\;c_{qg \rightarrow q}\;
\int_0^1 d z \left[
Q_i(x,\tau, t) \tilde G\left(\frac{x (1-z)}{z}\right)
\,\zeta\left(\frac{x}{z}, \tau \right)
\right.
\nonumber
\\
& & \;\;\;\;\;\;\;\;
\left.
\frac{}{}
\,-\,Q_i\left(x z, \tau, t \right)\, \tilde G\left( x (1-z)\right) \right]
\; \gamma_{q \rightarrow q g}(z)
\,\zeta\left(x, \tau \right)
\nonumber
\\
-\;
\hat B^\prime \{\tilde{\bar{Q}}_i\} \,Q_i &=&
-\;
\rho_N\;c_{q \bar q \rightarrow g}\;
\int_0^1 d z
\;Q_i(x,\tau,t) \tilde{\bar{Q}}_i\left(\frac{x (1-z)}{z} \right)
\; \gamma_{g \rightarrow q \bar q}(z)
\,\zeta\left(\frac{x}{z}, \tau \right)
\;\;.
\label{fq}
\end{eqnarray}
Here $\rho_N$ is the nucleon density of the nuclear medium (to be
specified below), and the constants
$c_{qg \rightarrow q} = c_{\bar q g \rightarrow \bar q}
= 1/8$, $c_{q\bar q\rightarrow g} = 8/9$ arise from the difference of
flux factors (color and spin) for fusions compared to branchings \cite{close}.
Similarly the fusion rates of antiquarks are given by replacing
$Q_i \leftrightarrow \bar Q_i$.
The corresponding loss and gain fusion rates of gluons are:
\begin{eqnarray}
-\;
\hat C^\prime \{\tilde G\} \, G &=&
-\;
\rho_N\;
c_{gg \rightarrow g}\;
\int_0^1 d z \bigg[
\,G(x,\tau,t) \tilde G \left( \frac{x (1-z)}{z} \right)
\,\zeta\left(\frac{x}{z}, \tau \right) \nonumber \\
&&\qquad \qquad -\,G\left(x z, \tau,t\right)\,
\tilde G\left( x (1-z) \right)
\,\zeta\left(x, \tau \right)
\bigg] \
\; \gamma_{g \rightarrow g g}(z)
\nonumber
\\
-\;
\hat D^\prime \{\tilde Q_j\} \,G &=&
-\;
\rho_N\;
c_{gq \rightarrow q}\;
G(x,\tau,t)\;
\int_0^1 d z
\; \tilde{Q}_j\left(\frac{x (1-z)}{z} \right)
\; \gamma_{q \rightarrow gq}(z)
\,\zeta\left(\frac{x}{z}, \tau \right)
\nonumber \\
-\;
\hat D^\prime \{\tilde{\bar Q}_j\} \,G &=&
-\;
\rho_N\;
c_{g q \rightarrow q}\;
G(x,\tau,t)\;
\int_0^1 d z
\; \tilde{\bar{Q}}_j\left(\frac{x (1-z)}{z} \right)
\; \gamma_{q \rightarrow gq}(z)
\,\zeta\left(\frac{x}{z}, \tau \right)
\nonumber
\\
+\;
\hat E^\prime \{\tilde{\bar{Q}}_j\} \, Q_j &=&
+\;
\frac{1}{2}\,
\rho_N\; c_{q \bar q \rightarrow g}\;
\int_0^1 d z
\; Q_j\left(x z\right) \tilde{\bar{Q}}_j \left(x (1-z) \right)
\; \gamma_{g \rightarrow q \bar q}(z)
\,\zeta\left(x, \tau \right)
\nonumber
\\
+\;
\hat E^\prime \{\tilde Q_j\} \, \bar Q_j &=&
+\;
\frac{1}{2}\,
\rho_N\; c_{q \bar q \rightarrow g}\;
\int_0^1 d z
\;\bar{Q}_j\left(x z\right) \tilde{Q}_j\left(x (1-z) \right)
\; \gamma_{g \rightarrow q \bar q}(z)
\,\zeta\left(x, \tau \right)
\;\;.
\label{fg}
\end{eqnarray}
The flux factors are here
$c_{g g \rightarrow g} = 1/8$,
$c_{g q \rightarrow q} = c_{g \bar q  \rightarrow \bar q}
= 1/8$, $c_{q\bar q\rightarrow g} = 8/9$.

In (\ref{fq}) and (\ref{fg}) the argument $x$ of the nuclear
parton distributions $\tilde Q$, $\tilde {\bar Q}$ and $\tilde G$
is understood as $\vert x \vert$, since the nuclear partons move in
$-P$ direction [c.f. eq. (\ref{e1})], but the measured
nucleon structure functions
are usually defined only for positive calues of $x$.
Also note that in (\ref{fg}) the second term on the right in
the expression for $\hat C^\prime \{\tilde G\} \,G$ would be multiplied
by a factor 1/2 if $\tilde G$ and $G$ would be treated on equal
footing, because nuclear and shower gluons would then be indistinguishable.
\smallskip

{\bf 2.4 Scattering processes}

\noindent
The $2 \rightarrow 2$ collision rates
for elastic scatterings of the cascading partons carrying momentum fractions
$x_1$ and $x_2$ with the partons in the nuclear background medium
of nucleon density $\rho_N$
carrying momentum fractions $x_1^\prime$ and $x_2^\prime$
are represented by the integral operator
\begin{eqnarray}
\hat S_{a b} \{\tilde A\} \, B  &\equiv&
-\;\int_{-1}^0 \frac{dx_2}{x_2} \;\tilde A(x_2) \;\int d p_\perp^2 \;
{d\hat\sigma_{ab\rightarrow ab}\over dp_\perp^2}
\;\rho_N
\;\frac{B(x,\tau,t)}{x}
\nonumber \\
&  & +
\frac{|x|P}{\tau^2}\;
\int_{-1}^0 \frac{dx_2}{x_2} \;\tilde A(x_2) \;
\left. {d\hat\sigma_{ab\rightarrow ab}
\over dp_\perp^2}  \right|_{p_\perp^2 = \frac{|x|P}{\tau}}
\nonumber \\
& & \;\;\;\;\;\;\;\;\;\;
\times \;
\;\rho_N
\int_0^\infty d \tau_1 \; \frac{B(x_1,\tau_1,t)}{x_1}
\;\left[1 \,+\theta\left(\frac{\tau_1}{\tau} - \frac{x_1}{\vert x\vert}
\right) \right]
\nonumber \\
&  &+
\;\int_{-1}^0 \frac{dx_2}{x_2} \;\tilde A(x_2)\;
\int^{x_1 P/\tau} d p_\perp^2\;
{d\hat\sigma_{ab\rightarrow ab}\over dp_\perp^2}
\;\rho_N
\;\frac{B(x_1,\tau,t)}{x_1}
\;\;,
\label{Sab}
\end{eqnarray}
where $A, B = G, Q_j, \bar Q_j$ and the ' $\tilde{} $ ' labels as before the
nuclear parton distributions, whereas the distributions without ' $\tilde{} $ '
refer to the cascading partons.
As in the case of fusions, eqs. (\ref{fq}) and (\ref{fg}),
the value of $x_2$ in the argument of the nuclear parton distribution
$\tilde A$ is meant as $\vert x_2 \vert$ when employing standard
parametrizations of nucleon structure functions.

The momentum fractions of two partons before $(x_1,x_2)$ and after
$(x_1^\prime, x_2^\prime)$ scattering involving a relative
transverse momentum exchange $p_\perp^2$ are related by
\begin{equation}
x_1\;=\; x_1^\prime \;+\;\frac{p_\perp^2}{(x_1^\prime - x_2) P^2}
\;\;\;,
\;\;\;\;\;\;\;\;\;
x_2\;=\; x_2^\prime \;+\;\frac{p_\perp^2}{(x_2^\prime - x_1) P^2}
\;\;,
\label{x12}
\end{equation}
from which follows the value of $x_1$ at which the function
$B(x_1,\tau,t)$ in eq. (\ref{Sab}) is to be evaluated,
\begin{equation}
x_1\;=\; x \;+\;\frac{p_\perp^2}{(x - x_2) P^2}
\;\;.
\label{x1}
\end{equation}

For the quarks (and similar antiquarks) one has the processes
$q_i g\rightarrow q_i g$, $q_i q_j \rightarrow q_i q_j$ and $q_i
\bar q_j \rightarrow q_i \bar q_j$,
whereas for gluons the contributing processes are
$g g\rightarrow g g$, $g q_j \rightarrow g q_j$ and $g \bar q_j
\rightarrow g \bar q_j$.
The parton-parton cross-sections $d \hat \sigma_{a b\rightarrow c d}/d
p_\perp^2$ in the scattering function (\ref{Sab})
are related to the squared scattering
amplitudes $\vert \overline{\cal M}_{a b \rightarrow c d}\vert ^2$, averaged
over initial spin and color states and summed over the final states,
by (neglecting the quark masses),
\begin{equation}
\frac{d \hat \sigma_{a b \rightarrow c d}(\hat s, p_\perp^2)}{d p_\perp^2}
\;=\;
{\cal D}_{ab}\,{\cal D}_{c d}\;
\frac{\pi \alpha_s^2(p_\perp^2)}{\hat s^2}\;\,
\vert \overline{\cal M}_{a b \rightarrow c d}\vert ^2
\;\;,
\label{dsig}
\end{equation}
where the degeneracy factors ${\cal D}_{ab} = (1+ \delta_{ab})^{-1}$
accounts for the identical particle effect in the initial state
if $a$ and $b$ are truly indistinguishable,
and correspondingly ${\cal D}_{cd}$ is the statistical factor for
the final state.  However, since cascading particles and nuclear
partons are assumed to be distinguishable, one has
${\cal D}_{ab} = 1$ always (but not so for ${\cal D}_{c d}$).
Since needed later, let me list for completeness
the relevant squared matrix elements for the various processes
\cite{fields}:
\begin{eqnarray}
\vert \overline{\cal M}_{q_i q_j \rightarrow q_i q_j}\vert^2 & = &
\frac{4}{9} \left(\frac{\hat{s}^2 + \hat{u}^2}{\hat{t}^2} \right) \;+\;
\delta_{i j} \; \left[ \frac{4}{9} \left(\frac{\hat{s}^2 +
\hat{t}^2}{\hat{u}^2} \right) \;-\;
\frac{8}{27} \left(\frac{\hat{s}^2}{\hat{u} \hat{t}} \right)
\right]
\nonumber
\\
\vert \overline{\cal M}_{q_i \bar q_j \rightarrow q_i \bar q_j}\vert^2 & = &
\frac{4}{9} \left(\frac{\hat{s}^2 + \hat{u}^2}{\hat{t}^2} \right) \;+\;
\delta_{i j} \; \left[ \frac{4}{9} \left(\frac{\hat{t}^2 +
\hat{u}^2}{\hat{s}^2} \right) \;-\;
\frac{8}{27} \left(\frac{\hat{u}^2}{\hat{s} \hat{t}} \right)
\right]
\nonumber
\\
\vert \overline{\cal M}_{q_i g \rightarrow q_i g}\vert^2 & = &
- \frac{4}{9} \left(\frac{\hat{u}^2 + \hat{s}^2}{\hat{u} \hat{s}} \right) \;+\;
 \left(\frac{\hat{u}^2 + \hat{s}^2}{\hat{t}^2} \right)
\nonumber
\\
\vert \overline{\cal M}_{g g \rightarrow g g}\vert^2 & = &
\frac{9}{2} \left(3 \;-\;\frac{\hat{u} \hat{t}}{\hat{s}^2} \;-\;
\frac{\hat{u} \hat{s}}{\hat{t}^2} \;-\;
\frac{\hat{s} \hat{t}}{\hat{u}^2} \right)
\;\;,
\label{M2}
\end{eqnarray}
The variables $\hat s$, $\hat t$, $\hat u$
are the kinematic invariants of the parton-parton scattering with
$\hat s + \hat t + \hat u = 0$, and $p_\perp^2 = \hat t \hat u / \hat s$
for massless particles.  For massive quarks the corresponding scattering
matrix-elements can be found in Ref. \cite{combridge}.
\bigskip

\noindent {\bf 3. ANALYTICAL SOLUTIONS FOR THE PARTON SHOWER FUNCTIONS}
\medskip

In the following I will present a method for solving the evolution
equations (\ref{eq})-(\ref{eg})
analytically and will investigate some characteristic features of
the solutions.
Recall the scenario from Sec. 2 with
a highly energetic incident quark or gluon,
produced at some point of time
$t_0$ inside the target nucleus,
with longitudinal momentum fraction $x_0 \simeq 1$,
transverse momentum $\vec k_{\perp 0}=\vec 0$,
and virtuality $Q_0^2 \gg \Lambda$, that initiates a cascade (or shower) of
secondary quarks, antiquarks, and gluons. By solving the evolution
equations, one can follow the time evolution of this parton shower
in the nuclear background medium consisting of the partons of the heavy
nucleus.
In the present section I will solve the evolution equations
(\ref{eq})-(\ref{eg})
for the parton momentum densities $Q$ ($\bar Q$) and $G$, or equivalently
the number densities $q$ ($\bar q$) and $g$.
The synonym {\it parton shower functions} as generic term for these
distributions will be frequently used.
As I will show, the evolution equations can be solved in closed form,
if one restricts to a somewhat idealized scenario
that allows for a transparent analysis
(without however
giving up the physical relevance of application to realistic situations).
\smallskip

\noindent
{\bf 3.1. Assumptions and approximations}
\smallskip

The assumptions and approximations that I will have to
impose in order to achieve tractable solutions are the following:
\begin{description}
\item[(i)]
As before I will employ a rigorous distinction  between the cascading partons
produced by the evolving parton shower and the partons of the nucleus which
represent the medium. Only the shower partons are
followed dynamically. The nuclear parton distributions are approximated
by a constant behaviour, motivated by a rather weak $x$-dependence of
the nucleon structure functions at small values of $x$ \cite{comment1}.
I will restrict the shower development to the small $x$-region
where the probability for an interaction of a parton is the largest,
because of the logarithmically large integration over the momenta of
the involved partons and the large available phase-space.
\item[(ii)]
The nuclear background medium is assumed to be
of very large (ideally of infinite)
size with a uniform spatial distribution of nuclear partons, so that
one can ignore boundary effects and average over the spatial coordinates.
Also, I make the so-called "sudden approximation", i.e. the nuclear
parton distributions remain essentially unaltered by the interactions
with the shower partons, at least on time scales
characteristic for the "intra-cascade activity". However, once a nuclear parton
has interacted with a shower parton, it is
considered to be materialized from the coherent medium and
becomes part of the evolving cascade. Thus the medium acts as
a particle and energy-momentum reservoir.
\item[(iii)]
For the nuclear partons of the medium,
I will ignore effects as nuclear shadowing, color coherence,
long range correlations, etc., that affect the momentum and number
distributions of those quanta. I approximate the nuclear distributions
as simple superposition of the experimentally measured nucleon structure
functions and consider the nuclear parton density as a function of
the number of nucleons $A$ per unit area.
The contributions of valence quarks are neglected, that is,
the matter is taken to be effectively baryon free, an assumption
that is reasonable for the valence quark depleted small $x$-region.
\end{description}

Accordingly, the attention is restricted here to the diffusion of
cascading partons in longitudinal direction and the
dissipation of longitudinal momentum (or energy) of the shower
particles. That is, I will solve for
the time evolution of the parton shower functions as a function
of longitudinal momentum fraction $x$ and age $\tau$.
A more realistic evolution including also the diffusion in
transverse momentum $\vec k_\perp$, the lateral spread in
$\vec r_\perp$, as well as equal treatment of shower
and nuclear partons,  requires a numerical analysis,
which will be addressed in a separate work.
\smallskip

\noindent
{\bf 3.2. Simplified form of the evolution equations}
\smallskip

Due to the dependence of the integral operators for the fusion processes
and the scattering processes on the parton distributions $\tilde Q$,
$\tilde{\bar{Q}}$, and $\tilde G$ of the nuclear background medium, the
right hand side of the evolution equations involves quadratic forms
of the parton distributions, because of the coupling of the shower partons
to the nuclear partons.
A general analytical solution of this non-linear problem
is therefore not possible.
To overcome this obstacle, I explicitely distinguish these two sources
[c.f. item (i) above]. I
approximate the nuclear parton densities as scale ($Q^2$) independent and
neglect their time variation, as well as nuclear shadowing effects, etc.,
[c.f. item (ii)] and simply represent them by
the nucleon structure functions multiplied with a nuclear factor,
\begin{equation}
\tilde Q_i(x) \;=\;
\rho_N\; x\, \tilde q_i(|x|, Q_0^2)
\;\;,\;\;\;\;\;\;\;\;\;
\tilde G(x) \;=\;
\rho_N\; x\, \tilde g(|x|, Q_0^2)
\;\;,
\label{pdx}
\end{equation}
where $\tilde q_i$, $\tilde g$ denote the usual (time-independent)
measured nucleon structure functions, in principle to be evaluated at
the initial scale $Q_0^2$ at which the primary parton is produced.
It is important to realize that in (\ref{pdx}) the values of
$x = p_z/P < 0$ in accordance with eq. (\ref{e1}), i.e. the fact that
the nuclear partons move in $-z$-direction. Therefore, although the
number densities $\tilde q_i$, $\tilde g$ are
(as probability densities) necessarily always
positive definite, the momentum densities are here $\tilde Q_i,
\tilde G < 0$.
The nuclear factor $\rho_N$ depends on the
geometry and the nucleon density \cite{close,mulqiu},
\begin{equation}
\rho_N\;=\;\frac{3}{2} \; R_A \, \bar n_N
\;\;,
\label{rhon}
\end{equation}
where $R_A= r_0\, A^{1/3}$, $r_0 = 1.2$ fm,  and
$\bar n_N = A / (4\pi R_A^3/3)$ are the nuclear radius and the
nucleon density of the traversed nucleus, respectively.
Furthermore, motivated by the aforementioned weak $x$-dependence
of the nucleon structure functions
at small values of
$x\,\lower3pt\hbox{$\buildrel < \over\sim$}\, 10^{-2}$ (see however
\cite{comment1}), one may
approximate the nuclear parton distributions (\ref{pdx}) by a
constant $\tilde Q^0$, respectively $\tilde G^0$,
times the nuclear factor (\ref{rhon}):
\begin{equation}
\tilde Q_i(x) \;=\;
\tilde{\bar Q}_i(x) \;=\;
-\;\frac{9}{8 \pi} \; \frac{A^{1/3}}{r_0^2}\; \eta_i \tilde Q^0\;\;,
\;\;\;\;\;
\tilde G(x) \;=\; -\;\frac{9}{8 \pi} \; \frac{A^{1/3}}{r_0^2}\; \tilde G^0
\;\;.
\label{pd0}
\end{equation}
Here the overall minus sign ensures that the nuclear momentum densities
are negative definite, as explained after eq. (\ref{pdx}).
These $x$- and $t$-independent parametrizations
mimic in a crude approximation the effective number of nuclear partons
per unit area, weighted with their longitudinal momentum,
as 'seen' by the evolving shower partons.
Since at small $x$ the contribution of the valence quarks is negligable,
the distributions $Q$ ($\bar Q$) are understood to represent solely
the sea quarks (sea antiquarks) for which I take
$n_f = 3$, that is, I account for the flavors
$i = 1, 2, 3 \equiv u, d, s$ with the relative proportions
\begin{equation}
\eta_i \;=\;\left\{ \begin{array}{rl}
\;1\;& \mbox{for $i = u \, (\bar u)$} \\
\;1\;& \mbox{for $i = d \, (\bar d)$} \\
\;0.43\;& \mbox{for $i = s \, (\bar s$)}
\end{array}
\right.
\;\;.
\label{etai}
\end{equation}
As reasonable values for
$\tilde Q^0$ and $\tilde G^0$ one may take:
\begin{equation}
\tilde Q^0 \;\approx\;0.2
\;\;,\;\;\;\;\;\;\;\;
\tilde G^0 \;\approx\;3
\;\;.
\end{equation}
As long one is concerned with color and flavor insensitive
kinetics of the parton evolution,
it is sufficient and convenient to consider $n_f=3$ identical
types of quarks and antiquarks and the only flavor dependence comes from
the relative quark admixtures (\ref{etai}) in the nuclear medium.
Due to the absence of valence quarks the baryon number is zero and
particle-antiparticle symmetry
implies $Q_i = \bar Q_i$. Therefore it is convenient to
define the flavor singlet combination by the sum over all quark flavors,
\begin{equation}
{\cal Q}\;:=\; \sum_{i=1}^{n_f}\;\left(\frac{}{} Q_i\; +\; \bar Q_i \right)
\label{qqb}
\end{equation}
and consider quarks and corresponding antiquarks as one particle species.

Collecting all these ingredients, one can now rewrite the evolution equations
(\ref{eq})-(\ref{eg}) in a much simpler, linear form:
\begin{eqnarray}
\frac{\partial}{\partial t} \, {\cal Q}(x, \tau, t)
&=&
-\;\left[
\,\hat A \,+ \,\hat A^\prime\{\tilde G^0\} \,+\,
\frac{1}{2} \,\hat B^\prime\{\kappa \tilde Q^0\} \,-\,
n_f \,\hat S_{qq}\{\kappa \tilde Q^0\}
\,+\, \frac{\partial}{\partial \tau}
\right] \; \,{\cal Q}
\nonumber \\
& &
+\;  \left[
\,2\,\hat B \,+ \,\hat S_{qg}\{\kappa \tilde Q^0\}
\right] \; \,G
\nonumber
\\
&\equiv&\, -\,\hat F_{qq} \; {\cal Q} \,\;+\;\, \hat F_{qg} \; G
\label{eq1}
\\
&&
\nonumber
\\
\frac{\partial}{\partial t} \, G(x, \tau, t)
&=&
-\; \left[
\,\hat C \,+\,  \hat D \,+\,
\hat C^\prime \{\tilde G^0\}\,+
\, n_f\,\hat D^\prime\{\kappa \tilde Q^0\} \,-\,
\hat S_{gg}\{\tilde G^0\}
\,+\, \frac{\partial}{\partial \tau}
\right] \; \,G
\nonumber \\
& &
+\; \left[
\,\hat E \,+\,  \frac{1}{2}\,\hat E^\prime \{\kappa \tilde Q^0\}\,+\,
\hat S_{gq}\{\tilde G^0\}
\right] \; \,{\cal Q}
\nonumber
\\
&\equiv&\, -\,\hat F_{gg} \; G \;\,+\;\, \hat F_{gq} \; {\cal Q}
\;\;.
\label{eg1}
\end{eqnarray}
Here $\eta_i$ is defined by (\ref{etai}) and the constant $\kappa$
accounts for the effective number of quark flavors in the nuclear medium
that are probed by the cascading gluons,
\begin{equation}
\kappa \;:=\; \frac{1}{n_f}\;\sum_{i=1}^{n_f} \,\eta_i \;=\;0.81
\;\;.
\end{equation}

In  (\ref{eq1}) and (\ref{eg1})
both the parton densities and the integral operators depend still
on the three variables $t$, $\tau$, and $x$, corresponding to
the time development of the vituality (age) distribution
and energy (longitudinal momentum) distribution, respectively.
For each age $\tau=|x|P/Q^2$ one must in general expect a different
$x$-dependence (and in principle also flavor dependence),
reflected by an "age distribution" ${\cal A}_a(\tau, t; x)$ which
represents the probability that a parton $a=q,g$ with $x$ at time $t$
has an age $\tau$, and which, for quarks and gluons, respectively,
is defined through \cite{gm1}
\begin{equation}
{\cal Q}(x,\tau,t) \;=\; {\cal A}_q(\tau, t; x)\,{\cal Q}(x,t)
\;\;, \;\;\;\;\;\;\;\;\;\;\;\;
G(x,\tau,t) \;=\; {\cal A}_g(\tau, t; x)\,G(x,t)
\;\;,
\label{agedist}
\end{equation}
with normalization $ \int_0^\infty d \tau {\cal A}_a(\tau, t; x)=1$,
and
\begin{equation}
{\cal A}_a(\tau,t)\;\equiv\; \int_0^1 dx\,{\cal A}_a(\tau,t;x)
\;\;,\;\;\;\;\;\;\;\;\;\;\;\;\;
\bar f(t)\;\equiv\;\int_0^\infty d \tau\int_0^1 dx \,{\cal A}_a(\tau,t;x)
\;f(x,\tau,t)
\;\;.
\label{ageprop}
\end{equation}
The age distribution evolves parallel with laboratory time in between
interactions (free streaming), but due to scatterings it receives modifications
because
the age of the scattered parton is reset to a younger age corresponding
to a larger virtuality caused by the momentum transfer of the interaction.
In the absence of a medium ${\cal A} = \delta(t-\tau)$.
Therefore the age distribution reflects the influence of the
nuclear medium and determines the characteristic age $\tau$. The age $\tau$
must hence be related to the mean free time in between scatterings and
consequently introduces an external scale depending on the density of the
medium.
The separation of the medium influence
in the parton densities by introducing ${\cal A}$, suggests to describe
the parton evolution
by the time rate of change with respect to the variable $x$ only and
interpret the age distribution  as externally induced feedback
to the shower functions via (\ref{agedist}) and (\ref{eq1}), (\ref{eg1}).
Therefore let me rewrite the latter evolution equations in terms of
${\cal Q}(x,t)$, $G(x,t)$ and ${\cal A}(\tau,t)$, defined in (\ref{ageprop}),
by substituting the factorized forms (\ref{agedist}):
\begin{eqnarray}
\frac{\partial}{\partial t} \, {\cal Q}(x, t)
&=&
\, -\,\left\{\,\hat F_{qq} \; +\;
\frac{1}{\cal A}\left(\frac{\partial}{\partial t}
+\frac{\partial}{\partial \tau}\right) {\cal A} \right\}\;{\cal Q}
\,\;+\;\, \hat F_{qg} \; G
\label{eq11}
\\
\frac{\partial}{\partial t} \, G(x, t)
&=&
\, \hat F_{gq} \; {\cal Q}
\, \;-\,\;\left\{\,\hat F_{gg} \; +\;
\frac{1}{\cal A}\left(\frac{\partial}{\partial t}
+\frac{\partial}{\partial \tau}\right) {\cal A} \right\}\;G
\;\;.
\label{eg11}
\end{eqnarray}
Here the terms involving ${\cal A}$ must in reality determined
selfconsistently by the dynamic interplay between the shower particles and the
nuclear environment. It contains implicitely information about the
nuclear structure which at this point has to be provided
as phenomenological input. However, the actual form of ${\cal A}$ is
irrelevant if one is not interested in the age distribution itself,
but rather is concerned with the time evolution of the
partons with respect to $x$ and {\it average} $\tau$.
In this case one may multiply both sides of the equations by
${\cal A}$, integrate over $\tau$ subject to the conditions
(\ref{ageprop}), so that the explicit dependence on
${\cal A}$ drops out. Then one is left with eqs. (\ref{eq11}), (\ref{eg11}),
now without the terms
${\cal A}^{-1}\left(\partial /\partial t
+\partial /\partial \tau\right) {\cal A}$
(taking ${\cal A}_q \approx {\cal A}_g$),
but one has to replace everywhere in the operators $\hat F_{ab}$
\begin{equation}
\tau \;\longrightarrow \;
\bar \tau (t) \;=\; v_\tau(t)\,t
\;\equiv\;
\int_0^\infty d\tau \;\tau\;{\cal A}(\tau,t)
\;\;,
\label{bartau}
\end{equation}
where
\begin{equation}
v_\tau(t)\;=\; \frac{d \bar \tau(t)}{dt}\;\, \le \;\,1
\label{vtau}
\end{equation}
is the "velocity" by which the
typical $\bar \tau$ changes with laboratory time $t$.
The evolution equations can now be written in a compact form
\cite{dokbook}:
If one represents the time evolving parton distributions
in a symbolic matrix form as $\hat U(t,t_0=0) =\exp[-i\hat H\,t]$
with $\hat H$
the Hamiltonian of the system, then the evolving parton state will
form a vector $({\cal Q}, G)$ which satisfies the
following Schr\"odinger type equation
\begin{equation}
\frac{\partial}{\partial t} \;
\left( \begin{array}{c} {\cal Q} \\ G \end{array} \right)
\;=\;
- i\;\hat H\;
\left( \begin{array}{c} {\cal Q} \\ G \end{array} \right)
\;=\;
\left( \begin{array}{rr}
- \hat F_{qq} \;& \hat F_{qg} \\
  \hat F_{gq} \;& - \hat F_{gg}
\end{array} \right) \;
\left( \begin{array}{c} {\cal Q} \\ G \end{array} \right)
\;\;.
\label{Schroe}
\end{equation}
As I will now show, this matrix equation, or equivalently
the simplified evolution equations (\ref{eq11}) and
(\ref{eg11}) averaged over $\tau$ according to (\ref{bartau}),
 can be solved in closed form by diagonalizing the
Hamiltonian $\hat H$, subject to some chosen
initial condition at $t=t_0$.
\medskip

\noindent
{\bf 3.3. Solutions for the parton evolution}
\smallskip

The integral operators in the matrix equation (\ref{Schroe})
are now forms of the average age
$\bar \tau(t)$ and the
longitudinal momentum fractions $x =p_z/P$ of the shower particles only.
As stated before, the time-dependent
$\bar \tau (t)$ loosely speaking reflects the
structure of the nuclear medium as probed by the shower partons,
and it reduces in free space to the trivial dependence $\bar \tau = t$,
i.e. $v_\tau =1$ in (\ref{vtau}).
Thus, one qualitatively expects a delayed parton evolution in medium
as compared to vacuum, because repeated rejuvenations of the
shower partons must lead to $\bar \tau < t$, or $v_\tau \le 1$.
If $v_\tau(t)$ has only a weak time-dependence, $v_\tau \approx const.$,
then an appropriate solution ansatz for the parton shower functions is
the following factorized form with a power dependence in $1/x$ and an
exponential decay with some time-dependent function $\chi(t)$:
\begin{eqnarray}
{\cal Q}(x, t) &=& \int_0^\infty d\tau {\cal A}_q(\tau, t) \;{\cal Q}(x,\tau,t)
\;=\;
{\cal N}_q \; \left(\frac{}{} \left|x\right|^{-s} \;\theta (x)\;+\;
\left|x\right|^{s} \;\theta (-x) \right)
\; \exp[- \mu\,\chi(t)]
\nonumber
\\
G(x, t) &=& \int_0^\infty d\tau {\cal A}_g(\tau, t) \;G(x,\tau,t)
\;=\;
{\cal N}_g \; \left( \frac{}{} \left|x\right|^{-s} \;\theta (x)\;+\;
\left|x\right|^{s} \;\theta (-x)\right)
\; \exp[- \mu\,\chi(t)]
\;\;.
\label{ansatz}
\end{eqnarray}
Let me explain the physical motivation of this ansatz:
\begin{description}
\item[(i)]
The introduced power $s$ is understood to be positive definite,
$s \ge 0$, and the constants  ${\cal N}_q$ and ${\cal N}_g$ are
normalization factors.
As will become clear, the 'parameter' $s$ plays the role of
a conjugate variable to the variable $x$, controlling the
longitudinal momentum distribution in $x$.
As the shower evolves, $s$ will change with time, thereby correlating the
$x$ and the $t$ dependence of the shower functions.
\item[(ii)]
The form of the $x$-dependence in brackets
is motivated by the particular dynamics viewed in the $CM_{NN}$ frame
(Sec. 2.1) in which one naturally expects some "drag effect" of the
(in $-z$ direction moving) nuclear partons on the evolving parton shower.
Although the shower partons initially
propagate in $+z$ direction, one
must account at some point for particles moving in both $+z$ and $-z$
direction. Shower partons moving in $+z$ direction will progressively
degrade their longitudinal momentum by branchings and scatterings, and
will eventually reverse their longitudinal direction, dragged along and
accelerated in $-z$ direction by absorption of (fusions with)
comoving nuclear partons.
\item[(iii)]
The exponential factor accounts for the fact that the running coupling
strength $\alpha_s$ in the operators $\hat F_{ab}$ of (\ref{Schroe})
depends on $\bar \tau(t)$ and therefore implicitely on $t$.
By introducing the function
\begin{equation}
\chi(t)\;\equiv\;\int_{t_0}^t d t'\,\frac{\alpha_s\left(\bar \tau(t')\right)}
{2 \pi \bar \tau(t')}
\;=\;\frac{1}{2 \pi b\,v_\tau}\;\ln\left(\frac{\alpha_s(v_\tau t)}
{\alpha_s(v_\tau t_0)}\right)
\;\;,
\label{chi}
\end{equation}
where $b=(11 n_c - 2 n_f)/12\pi = 0.72$ and $\bar \tau =v_\tau t$
with $v_\tau$ assumed to be only slowly varying with $t$,
it is possible to absorb the medium dependent factor
$\alpha_s(\bar \tau)/(2\pi \bar \tau)$
in the exponential time dependence of the distributions (\ref{ansatz}).
Note that for fixed coupling and $v_\tau =const.$, one has
$\chi \propto \ln(t/t_0)$. Even in the general case $\chi$ represents
approximately the logarithm of the laboratory time $t$.
The real time dependence in the laboratory frame is regained by the
inverse relation
\begin{equation}
t=\left(\frac{xP}{v_\tau \Lambda^2}\right)\;
\exp\left[-\,\ln\left(\frac{Q_0^2}{\Lambda^2}\right)
\exp(- 2 \pi b v_\tau \chi)\right]
\;\;,
\label{chi2}
\end{equation}
where $Q_0^2$ is the virtuality of the shower initiating parton at $t_0$.
\end{description}

\noindent
Substituting the ansatz (\ref{ansatz}) into the time-evolution
equation (\ref{Schroe}), one arrives at the following eigenvalue
equations with respect to the new time evolution
variable $\chi(t)$, corresponding
to the diagonalized Hamiltonian $\hat H$:
\begin{equation}
\frac{\partial}{\partial \chi} \;
\left( \begin{array}{c} {\cal Q} \\ G \end{array} \right)
\;=\;
- \,\mu\;
\left( \begin{array}{c} {\cal Q} \\ G \end{array} \right)
\;\;,
\label{eigenvalue}
\end{equation}
or equivalently
\begin{eqnarray}
\mu & = & F_{qq}(s) \;-\; \frac{b}{a}\; F_{qg}(s)
\nonumber
\\
\frac{b}{a} \; \mu & = & \frac{b}{a} \;F_{gg}(s) \;-\; F_{gq}(s)
\;\;.
\label{mu}
\end{eqnarray}
Here the scalar functions
$F_{ab}(s)$ correspond to the operators in the Hamiltonian, $\hat F_{ab}$,
defined by (\ref{eq1}) and (\ref{eg1}),
\begin{eqnarray}
F_{qq}(s) &=&
A(s) \;+\;A^\prime_{\{\tilde G^0\}}(s) \;+\;
\frac{1}{2}\,B^\prime_{\{\kappa \tilde Q^0\}}(s) \;-\;
n_f\,S^{(qq)}_{\{\kappa \tilde Q^0\}}(s)
\nonumber \\
F_{qg}(s) &=&
2 \;B(s) \;+\; S^{(qg)}_{\{\kappa \tilde Q^0\}}(s)
\nonumber \\
F_{gq}(s)&=&
\;n_f\,\hat E\;+\;
n_f\, E^\prime_{\{\kappa \tilde Q^0\}}(s) \;+\;
n_f\, S^{(gq)}_{\{\tilde G^0\}}(s)
\nonumber \\
F_{gg}(s)&=&
C(s) \;+\;C^\prime_{\{\tilde G^0\}}(s) \;+\;
D(s) \;+\;n_f\,D^\prime_{\{\kappa \tilde Q^0\}}(s) \;-\;
S^{(gg)}_{\{\tilde G^0\}}(s)
\;\;.
\label{fab}
\end{eqnarray}

The {\it branching kernels} in (\ref{fab}) are
\begin{eqnarray}
A(s) &=&
\int_0^1 dz \,\left(\frac{}{} 1 - z^{s} \right)
\;\gamma_{q\rightarrow qg}(z)
\;=\;
C_Q\;\left\{2 \,\psi (s+2) + 2 \gamma_{E}
- \frac{1}{(s+1)(s+2)} - \frac{3}{2} \right\}
\nonumber
\\
B(s) &=&
\int_0^1 dz \, z^{s} \;\gamma_{g\rightarrow q\bar q}(z)
\;=\;
\frac{1}{2} \;\left\{ \frac{s^2 + 3 s + 4}{(s+1)(s+2)(s+3)} \right\}
\nonumber
\\
C(s)\,+\,D(s) &=&
\int_0^1 dz \,\left(\frac{1}{2} - z^{s} \right)
\;\gamma_{g\rightarrow gg}(z)
\;+\;
n_f\;
\int_0^1 dz \, \gamma_{g\rightarrow q \bar q}(z)
\nonumber \\
&=&
C_G\;\left\{2 \,\psi (s+2) + 2 \gamma_{E}
\;-\; 4\,\frac{s^2 + 3 s + 3}{s(s+1)(s+2)(s+3)} \;-\;\frac{11}{6}
\;+\;\frac{n_f}{3 C_G} \right\}
\nonumber
\\
E(s) &=&
\int_0^1 dz \,z^{s}
\;\gamma_{q\rightarrow gq}(z)
\;=\;
C_Q\;\frac{s^2 +  3 s + 4}{s(s+1)(s+2)}
\;\;,
\label{bk}
\end{eqnarray}
where the branching functions $\gamma_{a\rightarrow bc}(z)$
are given by (\ref{gamma}), and as before $n_f=3$,
$C_Q=\frac{n_c^2-1}{2 n_c}$, $C_G=n_c$ with $n_c=3$.
The Psi (Digamma) function is defined as $\psi(z) = d[\ln \Gamma(z)]/dz$ and
$\gamma_E=0.5772$ is the Euler constant.
Note that these branching kernels are finite, i.e. free of infrared
singularities, because the infrared divergent contributions cancel
explicitely due to the balance of  gain and loss terms.

Similarly the {\it fusion kernels} are evaluated as:
\begin{eqnarray}
A^\prime_{\{\tilde G^0\}}(s) &=&
-\;\tilde G^0 \;h^A(v_\tau)\;c_{qg\rightarrow q}\;
\int_0^1 dz \,\left(\frac{}{} z - z ^{s} \right)
\;\gamma_{q\rightarrow qg}(z)
\nonumber \\
&=&
-\;\tilde G^0 \;h^A(v_\tau)\;c_{qg\rightarrow q}\;
C_Q\;\left\{\frac{}{}
2 \,\psi (s+2) + 2 \gamma_{E} - \frac{1}{(s+1)(s+2)} - \frac{17}{6}
\right\}
\nonumber
\\
B^\prime_{\{\kappa \tilde Q^0\}}(s) &=&
-\;\kappa \tilde Q^0 \;h^A(v_\tau)\;c_{q\bar q\rightarrow g}\;
\int_0^1 dz \, z \;\gamma_{g\rightarrow q\bar q}(z)
\;\,=\;\,
-\;\kappa \tilde Q^0 \;h^A(v_\tau)\;c_{q\bar q\rightarrow g}\;
\frac{1}{6}
\nonumber \\
C^\prime_{\{\tilde G^0\}}(s) + n_f\,D^\prime_{\{\kappa \tilde Q^0\}}(s) &=&
-\;\tilde G^0 \;h^A(v_\tau)\;c_{g g\rightarrow g}\;
\int_0^1 dz \,\left(\frac{}{} z - z^{s} \right)
\;\gamma_{g\rightarrow gg}(z)
\nonumber \\ & &
\;-\;
\kappa \tilde Q^0 \;h^A(v_\tau)\;c_{g q\rightarrow q}\;
n_f\;\int_0^1 dz \, z \;\gamma_{q\rightarrow g q}(z)
\nonumber \\
&=&
-\;\tilde G^0 \;h^A(v_\tau)\;c_{g g\rightarrow g}\;
C_G\;\left\{\frac{}{}
2 \,\psi (s+2) + 2 \gamma_{E} - 4 \frac{s^2+3s+3}{s(s+1)(s+2)(s+3)}
\right.
\nonumber \\ & &
\left.
- \;\frac{11}{6}
\;+\;\frac{\kappa \tilde Q^0}{\tilde G^0} \,
\frac{c_{gq\rightarrow q}}{c_{gg\rightarrow g}}\,
\frac{4 C_Q}{3 C_G}\,n_f \right\}
\nonumber
\\
E^\prime_{\{\kappa \tilde Q^0\}}(s) &=&
-\;\kappa \tilde Q^0 \;h^A(v_\tau)\;c_{q \bar q\rightarrow g}\;
\int_0^1 dz \,z^{s}
\;\gamma_{g\rightarrow q\bar q}(z)
\nonumber \\
&=&
-\;\kappa \tilde Q^0 \;h^A(v_\tau)\;c_{q \bar q\rightarrow g}\;
\frac{1}{2} \; \frac{s^2 + 3 s +4}{(s+1)(s+2)(s+3)}
\;\;.
\label{fk}
\end{eqnarray}
The constants $c_{ab\rightarrow c}$ in (\ref{fk}) are given after eqs.
(\ref{fq}) and (\ref{fg}).
The dimensionless function $h^A$ contains the nuclear dependence,
eqs. (\ref{rhon}) and (\ref{pd0}),
\begin{eqnarray}
h^A(v_\tau)&=& \frac{9 \pi\,A^{1/3}}{r_0^2}\;\frac{v_\tau t}{|x|P}
\nonumber
\\
&=& \frac{9 \pi\,A^{1/3}}{r_0^2 \Lambda^2}\;
\exp\left[-\,\ln\left(\frac{Q_0^2}{\Lambda^2}\right)
\exp(- 2 \pi b v_\tau \chi)\right]
\;\;.
\label{htau}
\end{eqnarray}

Finally the {\it scattering kernels} are obtained as:
\begin{eqnarray}
S^{(qq)}_{\{\kappa \tilde Q^0\}}(s) &=&
\frac{1}{4}\;\alpha_s(v_\tau t)\;\kappa \tilde Q^0\; h^A(v_\tau)\;
\Sigma_{qq\rightarrow qq}
\;\;\;\;\;\;\;\;\;
S^{(qg)}_{\{\kappa \tilde Q^0\}}(s) \;=\;
\frac{1}{4}\;\alpha_s(v_\tau t)\;\kappa \tilde Q^0\; h^A(v_\tau)\;
\Sigma_{qg\rightarrow qg}
\nonumber \\
S^{(gq)}_{\{\tilde G^0\}}(s) &=&
\frac{1}{4}\;\alpha_s(v_\tau t)\; \tilde G^0\; h^A(v_\tau)\;
\Sigma_{gq\rightarrow gq}
\;\;\;\;\;\;\;\;\;\;\;\;
S^{(gg)}_{\{\tilde G^0\}}(s) \;=\;
\frac{1}{4}\;\alpha_s(v_\tau t)\; \tilde G^0\; h^A(v_\tau)\;
\Sigma_{gg\rightarrow gg}
\;\;.
\label{sk}
\end{eqnarray}
These scattering kernels result from the operators (\ref{Sab})
in the limit of the "small-angle approximation" or "Landau approximation"
\cite{scott}, in which one restricts to scatterings with
small transverse momentum exchange $p_\perp^2 \ll \hat s < P^2$.
In this case the scattering operator (\ref{Sab}) simplifies
considerably to
\begin{eqnarray}
\hat S_{a b} \{\tilde A\} \, B  &\approx&
\int_{0}^{\infty} d \tau_1 \int_0^1 \frac{dx_1}{x_1}
\int_{-1}^0 \frac{dx_2}{x_2} \int_0^\infty dp_\perp^2
B(x_1,\tau,t) \,\tilde A(x_2)
\nonumber \\
& & \;\;\times \;\rho_N\;
\left.
{d\hat\sigma_{ab\rightarrow ab} \over dp_\perp^2}
\right|_{p_\perp^2 = \frac{|x|P}{\tau}}
\;\delta\left( x_2^\prime -x \right)
\,\delta\left(\tau - \frac{\vert x \vert P}{p_\perp^2} \right)
\;\;.
\label{Sab1}
\end{eqnarray}
where scattering at small angles implies [c.f. eq. (\ref{x12})]
\begin{equation}
x_1\;=\; x_1^\prime \;+\;\frac{p_\perp^2}{(x_1^\prime - x_2) P^2}
\;\approx \; x_1^\prime
\;\;\;,
\;\;\;\;\;\;\;\;\;
x_2\;=\; x_2^\prime \;+\;\frac{p_\perp^2}{(x_2^\prime - x_1) P^2}
\;\approx \; x_2^\prime
\;\;,
\end{equation}
and the parton-parton cross-sections
$d \hat \sigma_{ab\rightarrow ab}/d p_\perp^2$ given by (\ref{dsig}) and
(\ref{M2}) can be replaced by their small-angle contributions
\begin{equation}
\frac{d \hat \sigma_{ab\rightarrow ab}}{d p_\perp^2}\;\,\approx \;\,
\frac{\pi \alpha_s^2}{p_\perp^4} \; \Sigma_{ab \rightarrow ab}
\;\;,
\end{equation}
where the constants $\Sigma_{ab\rightarrow ab}$
follow from the squared matrix elements
$\vert \overline{\cal M}_{ab\rightarrow ab}\vert^2$, eq. (\ref{M2}),
by observing that at small angles $p_\perp^2 = - \hat t \ll \hat s$
and keeping only the leading terms in $p_\perp^2/\hat s$:
\begin{eqnarray}
\Sigma_{qq\rightarrow qq} &\equiv&
\Sigma_{q_i q_j\rightarrow q_i q_j}=
\Sigma_{q_i \bar q_j\rightarrow q_i \bar q_j} =
\Sigma_{q_i q_i\rightarrow q_i q_i}=
\Sigma_{\bar q_i \bar q_i\rightarrow \bar q_i \bar q_i}=
\Sigma_{q_i \bar q_i\rightarrow q_i \bar q_i}\;=\; \frac{8}{9}
\nonumber\\
\Sigma_{qg\rightarrow qg} &\equiv&
\Sigma_{q_i g\rightarrow q_i g}=
\Sigma_{\bar q_i g\rightarrow \bar q_i g}\;=\; 2
\nonumber\\
\Sigma_{gq\rightarrow gq} &\equiv&
\Sigma_{g q_i\rightarrow g q_i}=
\Sigma_{\bar g q_i \rightarrow g \bar q_i}\;=\; 2
\nonumber \\
\Sigma_{g g\rightarrow g g}&=& \frac{9}{2}
\;\;.
\end{eqnarray}

Returning to the eigenvalue equations (\ref{mu}) and
combining them, one obtains
\begin{equation}
\left[\frac{}{} \mu \;+\; \lambda_+(s)\right] \;
\left[\frac{}{} \mu \;+\; \lambda_-(s)\right] \;=\;0
\;\;,
\end{equation}
which implies
\begin{eqnarray}
\lambda_+(s) &=& - \frac{1}{2}\, \left[ \frac{}{}
F_{qq}(s) \; + \; F_{gg}(s) \right]\; +\;
\frac{1}{2}\,\sqrt{\left[F_{qq}(s) \,-\,F_{gg}(s)\right]^2
\;+\; 4\, F_{qg}(s)\,F_{gq}(s) }
\nonumber \\
\lambda_-(s) &=& - \frac{1}{2}\, \left[ \frac{}{}
F_{qq}(s) \; + \; F_{gg}(s) \right]\; -\;
\frac{1}{2}\,\sqrt{\left[F_{qq}(s) \,-\,F_{gg}(s)\right]^2
\;+\; 4\, F_{qg}(s)\,F_{gq}(s) }
\;\;,
\label{lambda+-}
\end{eqnarray}
and, corresponding to these functions, the constraints
\begin{eqnarray}
\frac{{\cal N}_g}{{\cal N}_q} &=& \frac{F_{gq}(s)}{F_{gg}(s)\,+\,\lambda_+(s)}
\;\;,\;\;\;\;\;\;\;\; \hbox{for} \;\;\;
\mu \;=\; - \lambda_+
\;\;,
\end{eqnarray}
or
\begin{eqnarray}
\frac{{\cal N}_g}{{\cal N}_q} &=& \frac{F_{gq}(s)}{F_{gg}(s)\,+\,\lambda_-(s)}
\;\;,\;\;\;\;\;\;\;\; \hbox{for} \;\;\;
\mu \;=\; - \lambda_-
\;\;.
\end{eqnarray}
Thus the solutions are given by
\begin{eqnarray}
{\cal Q}(x, t) &=&  {\cal N}_q \,\left(\frac{1}{x}\right)^{s}\;
\exp[ \lambda_+(s) \,\chi(t)]
\nonumber \\
G(x, t) &=&  \frac{{\cal N}_q\,F_{gq}(s)}{F_{gg}(s)\,+\,\lambda_+(s)}
 \,\left(\frac{1}{x}\right)^{s}\;
\exp[ \lambda_+(s) \,\chi(t)]
\;\;,
\end{eqnarray}
or
\begin{eqnarray}
{\cal Q}(x, t) &=&  {\cal N}_q \,\left(\frac{1}{x}\right)^{s}\;
\exp[ \lambda_-(s) \,\chi(t)]
\nonumber \\
G(x, t) &=&  \frac{{\cal N}_q\,F_{gq}(s)}{F_{gg}(s)\,+\,\lambda_-(s)}
 \,\left(\frac{1}{x}\right)^{s}\;
\exp[ \lambda_-(s) \,\chi(t)]
\;\;,
\end{eqnarray}
or by a linear combination of these two solutions. In general one can write the
solutions for the parton number distributions
$\wp = {\cal Q}/x = \sum_i \left( q_i + \bar q_i\right)$ and $g = G/x$
as:
\begin{eqnarray}
\wp(x, t) dx &=&  \frac{d x}{x^{s+1}}\;
\left[ \frac{}{} a_+\; e^{\lambda_+(s)\,\chi(t)}
\;+\; a_- \; e^{\lambda_-(s)\, \chi(t)} \right]
\nonumber
\\
g(x, t) dx  &=&  \frac{d x}{x^{s+1}}\;
\left[ \frac{a_+ \, F_{gq}}{F_{gg}\,+\,\lambda_+}\; e^{\lambda_+(s)\,\chi(t)}
\;+\; \frac{a_- \, F_{gq}}{F_{gg}\,+\,\lambda_-}\;
e^{\lambda_-(s)\, \chi(t)} \right]
\;\;.
\label{sol1}
\end{eqnarray}

Fig. 2 shows the behaviour of the functions
$F_{qq}, F_{qg}, F_{gq}, F_{gg}$ and Fig. 3 of
$\lambda_+, \lambda_-$,
and the derivatives $\lambda_+^\prime$, $\lambda_+^{\prime\prime}$
with respect to the variable $s$.
The three cases a)-c) correspond to different nuclear density of the medium
as controlled by the nuclear mass number entering the
function $h^A$, eq. (\ref{htau}): a) $A=0$
(free space), b) $A=50$ (moderately dense medium),
and c) $A=200$ (dense medium).
Evidently $\lambda_+$  vanishes at $s=s_0 \simeq 1.0; 1.1; 1.2$ for
the cases a); b); c), respectively,
and is positive
for $s < s_0$, and negative
for $s > s_0$. From (\ref{sol1}) it is obvious that
the number distributions of shower partons
behave as $dx / x^{s+1}$, so that one can
interpret the region with $s < s_0$ as the developing
stage of the shower, with $s = s_0$ representing the
situation at the shower maximum,
and the region with $s > s_0$ as the fading stage leading to the tail end of
the shower.
The variable $s$ therefore indicates the shower age at a given value of $x$: in
a
`young' shower ($s < s_0$) the number of shower particles increases with age,
while in an `old'
shower ($s > s_0$), it decreases.
\smallskip

Solutions corresponding to a certain initial
condition are obtained by summing up these
parton shower functions with respect to the
variable $s$, weighting them with an
appropriate amplitude. This procedure is known
as the Mellin transformation method.
It has been applied extensively to similar problems
of electron-photon showers in the
early days of cosmic ray theory \cite{cr1,cr2}.
The reason for introducing Mellin transforms of the parton shower functions
is that they are more easily determined than the original functions.
Let me denote by ${\sl M}_q$ (${\sl M}_g$) the Mellin transform of the function
$\wp$ ($g$),
\begin{equation}
{\sl M}_q (s,t)\;=\; \int_0^1 d x \, x^s\; \wp(x,t)
\;\;\;,\;\;\;\;\;\;\;
{\sl M}_g (s,t)\;=\; \int_0^1 d x \, x^s\; g(x,t)
\;\;,
\label{Mellin}
\end{equation}
then the inverse transformation is
\begin{equation}
\wp(x,t)\;=\; \frac{1}{2 \pi i}
\int_{c-i\infty}^{c+i\infty} d s \frac{{\sl M}_q (s,t)}{x^{s+1}}
\;\;\; , \;\;\;\;\;\;
g(x,t)\;=\; \frac{1}{2 \pi i} \int_{c-i\infty}^{c+i\infty} d s
\frac{{\sl M}_g (s,t)}{x^{s+1}}
\;\;,
\end{equation}
where $s$ is now a complex parameter,
and the path of the integral is running paralell to the imaginary axis.
By analytic continuation of the shower functions
(\ref{sol1}) to complex values of $s$, one obtains the solutions
\begin{eqnarray}
\wp(x, t)&=& \frac{1}{2 \pi i} \int_{c-i\infty}^{c+i\infty} d s
\frac{1}{x^{s+1}}
\;\left[ \frac{}{} a_+\; e^{\lambda_+(s)\,\chi(t)}
\;+\; a_- \; e^{\lambda_-(s)\, \chi(t)} \right]
\nonumber
\\
g(x, t)&=& \frac{1}{2 \pi i}
\int_{c-i\infty}^{c+i\infty} d s \frac{1}{x^{s+1}}
\;\left[ \frac{a_+ \, F_{gq}}{F_{gg}\,+\,\lambda_+}\; e^{\lambda_+(s)\,\chi(t)}
\;+\; \frac{a_- \, F_{gq}}{F_{gg}\,+\,\lambda_-}\;
e^{\lambda_-(s)\, \chi(t)} \right]
\;\;,
\label{sol2}
\end{eqnarray}
where $a_+$ and $a_-$ are functions of the complex parameter $s$, and are
determined by the initial
conditions imposed on $\wp$ and $g$.

I will now discuss the two cases of
1) a parton shower initiated by a quark, and 2)
a parton shower triggered by a primary gluon.
In accord with the physical picture that was outlined in Sec. 2,
the shower axis $z$ is defined along the momentum of the primary quark or
gluon,
so that the incident parton has zero transverse momentum. Since I consider here
the longitudinal evolution only,
the time development of the parton cascade proceeds by increasingly
populating the region with small values of $x$ as the shower
develops along the shower axis which corresponds to growing values of $s$.
According to eq. (\ref{p0}), the initiating primary parton carries a
longitudinal momentum
$p_{z 0} = x_0 P$ and has a initial time-like virtuality
$Q_0^2$ such that its energy is
determined by $E_0 = \sqrt{(x_0 P)^2 + Q_0^2} = x_0 P+ Q_0^2/(2x_0 P)$,
where $Q_0 \ll P$ is assumed.
%\smallskip

\newpage

\noindent
{\bf 3.3.1  Parton shower initiated by an incident quark}

If one considers a parton shower initiated by a primary quark,
then the initial conditions
at time $t=t_0$ are
\begin{equation}
\wp (x, t_0) \;=\; \delta(x_0 - x) \;
\delta\left(E_0-P\sqrt{x_0^2+\frac{Q_0^2}{P^2}}\right)
\;\;,
\;\;\;\;\;\;\;
g(x, t_0) \;=\; 0
\;\;.
\label{sol3}
\end{equation}
Inserting this initial condition into (\ref{Mellin}), one has at $t_0=0$,
\begin{equation}
{\sl M}_q(s,t_0) \;=\; x_0^s \;\;, \;\;\;\;\;\;\; {\sl M}_g (s,t_0)\;=\; 0
\;\;,
\label{sol4}
\end{equation}
which means that
\begin{equation}
\wp(x, t=0) \;=\; \frac{1}{2 \pi i} \int_{c-i\infty}^{c+i\infty} d s
\left(\frac{x_0}{x}\right)^s
\,\frac{1}{x}
\;\;,\;\;\;\;\;\;\;
g(x, t=0) \;=\; 0
\;\;.
\label{sol5}
\end{equation}
The parameters $a_+$ and $a_-$ are now completely determined by
(\ref{sol3})-(\ref{sol5}), together with (\ref{sol2}),
\begin{equation}
a_+ \;+\;a_- \;=\; x_0^s \;\;,\;\;\;\;\;\;\;
\frac{a_+ \, F_{gq}}{F_{gg}\,+\,\lambda_+}\;+\;
\frac{a_- \, F_{gq}}{F_{gg}\,+\,\lambda_-}\; =\; 0 \;\;.
\end{equation}
If one introduces functions $H_+$ and $H_-$ through
\begin{equation}
a_+ \;=\; H_+(s)\; x_0^s \;\;,\;\;\;\;\;\;\; a_- \;=\; H_-(s)\; x_0^s \;\;,
\end{equation}
or
\begin{equation}
H_+(s) \;=\;\frac{F_{gg}(s) \;+\;
\lambda_+(s)}{\lambda_+(s) \;-\;\lambda_-(s)} \;\;,\;\;\;\;\;\;\;
H_-(s) \;=\;-\,\frac{F_{gg}(s) \;+\;
\lambda_-(s)}{\lambda_+(s) \;-\;\lambda_-(s)}
\;\;,
\end{equation}
one can now express the solutions for the {\it differential spectra} of
secondary quarks and gluons
produced by a primary quark as
\begin{eqnarray}
\wp(x, t) dx&=& \frac{1}{2 \pi i} \;\frac{dx}{x}\;\int_{c-i\infty}^{c+i\infty}
d s \left(\frac{x_0}{x}\right)^s
\;\left[ \frac{}{} H_+(s)\; e^{\lambda_+(s)\,\chi(t)}
\;+\; H_-(s) \; e^{\lambda_-(s)\, \chi(t)} \right]
\nonumber
\\
g(x, t) dx&=& \frac{1}{2 \pi i} \;\frac{dx}{x}\;\int_{c-i\infty}^{c+i\infty} d
s \left(\frac{x_0}{x}\right)^s
\;\frac{L(s)}{\sqrt{s}} \;\left[ \frac{}{} e^{\lambda_+(s)\,\chi(t)}
\;-\; e^{\lambda_-(s)\, \chi(t)} \right]
\;\;,
\label{diffq}
\end{eqnarray}
where
\begin{equation}
L(s) \;=\; \frac{\sqrt{s}\, F_{gq}(s)}{\lambda_+(s) \;-\;\lambda_-(s)} \;\;.
\end{equation}
Correspondingly, one can evaluate the {\it integral spectra}
of secondary quarks and gluons, that is the integrated number densities
of partons with longitudinal momentum fractions greater than a given
value $x$,
\begin{eqnarray}
I_q(x, t)&=&
\int_{x}^{x_0} d x^\prime \, q(x^\prime, p_\perp^2, t) \;=\;
\frac{1}{2 \pi i} \int_{c-i\infty}^{c+i\infty} \frac{d s}{s}
\left(\frac{x_0}{x}\right)^s
\;\left[ \frac{}{} H_+(s)\; e^{\lambda_+(s)\,\chi(t)}
\;+\; H_-(s) \; e^{\lambda_-(s)\, \chi(t)} \right]
\nonumber
\\
I_g(x, t)&=&
\int_{x}^{x_0} d x^\prime \, g(x^\prime, p_\perp^2, t) \;=\;
\frac{1}{2 \pi i} \int_{c-i\infty}^{c+i\infty} \frac{d s}{s^{3/2}}
\left(\frac{x_0}{x}\right)^s
\;L(s) \;\left[ \frac{}{} e^{\lambda_+(s)\,\chi(t)}
\;-\; e^{\lambda_-(s)\, \chi(t)} \right]
\;\;.
\label{intq}
\end{eqnarray}
%\smallskip

\noindent
{\bf 3.3.2 Parton shower initiated by an incident gluon}

If one considers instead of a primary quark an incident gluon, the initial
conditions are
\begin{equation}
\wp(x, t_0) \;=\; 0
\;\;,
\;\;\;\;\;\;\;
g(x, t_0) \;=\; \delta(x_0 - x)
\;\delta\left(E_0-P\sqrt{x_0^2+\frac{Q_0^2}{P^2}}\right)
\;\;,
\end{equation}
the procedure is completely analogous. In this case one has instead of
(\ref{diffq}) for the
{\it differential spectra}:
\begin{eqnarray}
\wp(x, t) dx&=& \frac{1}{2 \pi i} \;\frac{dx}{x}\;\int_{c-i\infty}^{c+i\infty}
d s \left(\frac{x_0}{x}\right)^s
\;\sqrt{s}\,J(s) \;\left[ \frac{}{} e^{\lambda_+(s)\,\chi(t)}
\;-\; e^{\lambda_-(s)\, \chi(t)} \right]
\nonumber
\\
g(x, t) dx&=& \frac{1}{2 \pi i} \;\frac{dx}{x}\;\int_{c-i\infty}^{c+i\infty} d
s \left(\frac{x_0}{x}\right)^s
\;\left[ \frac{}{} H_-(s)\; e^{\lambda_+(s)\,\chi(t)}
\;+\; H_+(s) \; e^{\lambda_-(s)\, \chi(t)} \right]
\;\;,
\label{diffg}
\end{eqnarray}
where
\begin{equation}
J(s) \;=\; \frac{1}{\sqrt{s}} \frac{F_{qg}(s)}{\lambda_+(s) \;-\;\lambda_-(s)}
\;\;.
\end{equation}
For the {\it integral spectra} one has instead of (\ref{intq}):
\begin{eqnarray}
I_q(x, t)&=& \frac{1}{2 \pi i} \int_{c-i\infty}^{c+i\infty} \frac{d
s}{\sqrt{s}} \left(\frac{x_0}{x}\right)^s
\;J(s) \;\left[ \frac{}{} e^{\lambda_+(s)\,\chi(t)}
\;-\; e^{\lambda_-(s)\, \chi(t)} \right]
\nonumber
\\
I_g(x, t)&=& \frac{1}{2 \pi i} \int_{c-i\infty}^{c+i\infty} \frac{d s}{s}
\left(\frac{x_0}{x}\right)^s
\;\left[ \frac{}{} H_-(s)\; e^{\lambda_+(s)\,\chi(t)}
\;+\; H_+(s) \; e^{\lambda_-(s)\, \chi(t)} \right]
\;\;.
\label{intg}
\end{eqnarray}
%\smallskip

\noindent
{\bf 3.3.3 Saddle point evaluation of the parton spectra}

The integrals $F=\wp, g, I_q, I_g$ can be evaluated
by numerical computation of the contour integral in the $s$ plane,
but alternatively one can also obtain analytical forms by
employing the saddle point method which I will outline now.
First note that Fig. 3  exhibits the fact that $\lambda_+ > \lambda_-$ for
all values of $s$, so that one can neglect the second term
in these integrals
($\propto \exp[\lambda_- \,\chi]$) compared to the first term
($\propto \exp[\lambda_+ \,\chi]$).
Therefore, dropping the second term the typical structure
of the integrals is of the form
\begin{eqnarray}
F(x,t)&=&
\frac{1}{2 \pi i} \int_{c-i\infty}^{c+i\infty} d s \, s^{-k}\,
\left(\frac{x_0}{x}\right)^s {\cal H}(s)\;
e^{\lambda_+(s)\,\chi(t)}
\nonumber
\\
&=&
\frac{1}{2 \pi i} \int_{c-i\infty}^{c+i \infty}
d s \, {\cal H}(s)\,\exp[\lambda_+(s) \, \chi \,+\, y s \,- \, k \ln s]
\;\;,
\label{F}
\end{eqnarray}
where the symbol ${\cal H}$ under the integral
stands for the functions $H_+, H_-, L, J$,
the variable $y=\ln(x_0/x)$, and
the values of the power $k$ are  -1/2, 0, 1/2, 1, or 3/2, as can be read off
the formulae of Secs. 3.3.1 and 3.3.2.
Now, the functions  ${\cal H}$ are slowly varying with $s$, so that
the integrand in (\ref{F}) is a product of a function
${\cal H}$ with a weak $s$-dependence times an exponential of
the form $\exp[\phi(s)]$, where
\begin{equation}
\phi(s) \;=\;  \lambda_+(s) \, \chi \,+\, y s \,- \, k \ln s
\;\;.
\label{phi}
\end{equation}
If $\partial \phi/\partial s =0$
and $\partial ^2 \phi/\partial s^2 > 0$, at  some point $s = \bar s$,
then the integrand in (\ref{F}) may be approximated
by a Gaussian function on the path
parallel to the imaginary axis crossing the real axis at
$s = \bar s$, with the slowly varying
functions ${\cal H}(s)$ treated as constants and
evaluated at $\bar s$. Thus, by expanding $\phi(s)$ around $s = \bar s$ where
$d \phi/d s = 0$, one can evaluate the integrals
by the saddle point method (see e.g. \cite{cr1}),
\begin{equation}
F(x,t)\;=\;\left(2 \pi \left.\frac{\partial \phi}{\partial s}\right|_{s=\bar s}
\right)^{-1/2}\;\left(\frac{x_0}{x}\right)^{\bar s}\;\exp[\lambda_+(\bar s)
\chi]
\;\;,
\end{equation}
with the {\it saddle point} $\bar s$
defined by the equation
\begin{equation}
\frac{\partial \phi}{\partial s} \;=\;
\lambda_+^\prime (\bar s) \, t \,+\, y \,- \, \frac{k}{\bar s} \;=\;0
\label{saddle}
\;\;,
\end{equation}
which establishes a correlation between $y=\ln(x_0/x)$ and $\chi$,
i.e. between the momentum distribution and the time evolution variable $\chi$.
Applying this formalism, one obtains the following results
for the parton spectra $F=\wp, g, I_q, I_g$
(dropping the bar on the $s$, i.e. $s \equiv \bar s$ is the saddle point):

1) For a primary quark of longitudinal momentum fraction $x_0$,
\begin{eqnarray}
\wp(x, t) \, d x&=&  \frac{1}{\sqrt{2 \pi}}\,
\frac{H_+(s)}{\sqrt{\lambda_+^{\prime\prime}(s) \, \chi}}
\left(\frac{x_0}{x}\right)^s \; \frac{d x}{x}\;e^{\lambda_+(s) \,\chi}
\;,\;\;\;\;\;\;
\chi(t) \;=\; -\frac{1}{\lambda_+^\prime(s)}\,\ln\left(\frac{x_0}{x}\right)
\nonumber
\\
g(x, t) \, dx&=&  \frac{1}{\sqrt{2 \pi\, s}}\,
\frac{L(s)}{\sqrt{\lambda_+^{\prime\prime}(s) \, \chi\,+\,1/(2 s^2)}}
\left(\frac{x_0}{x}\right)^s \; \frac{d x}{x}\;e^{\lambda_+(s) \,\chi}
\;,\;\;\;\;\;\;
\chi(t)\;=\;
-\frac{1}{\lambda_+^\prime(s)}\,\left[\ln\left(\frac{x_0}{x}\right)
\;-\;\frac{1}{2 s}\right]
\nonumber
\\
I_q(x, t) &=&  \frac{1}{\sqrt{2 \pi\, s^2}}\,
\frac{H_+(s)}{\sqrt{\lambda_+^{\prime\prime}(s) \, \chi\,+\,1/s^2}}
\left(\frac{x_0}{x}\right)^s \; e^{\lambda_+(s) \,\chi}\;,\;\;\;\;\;\;
\chi(t) \;=\; -\frac{1}{\lambda_+^\prime(s)}\,
\left[\ln\left(\frac{x_0}{x}\right)\;-\;\frac{1}{s}\right]
\nonumber
\\
I_g(x, t) &=&  \frac{1}{\sqrt{2 \pi\, s^3}}\,
\frac{L(s)}{\sqrt{\lambda_+^{\prime\prime}(s) \, \chi\,+\,3/(2s^2)}}
\left(\frac{x_0}{x}\right)^s \; e^{\lambda_+(s) \,\chi}\;,\;\;\;\;\;\;
\chi(t) \;=\; -\frac{1}{\lambda_+^\prime(s)}\,
\left[\ln\left(\frac{x_0}{x}\right)
\;-\;\frac{3}{2 s}\right]
\;\;.
\label{qspec}
\end{eqnarray}

2) For a primary gluon of longitudinal momentum fraction $x_0$,
\begin{eqnarray}
\wp(x, t) \, d x&=&  \sqrt{\frac{s}{2 \pi}}\,
\frac{J(s)}{\sqrt{\lambda_+^{\prime\prime}(s) \, \chi\,-\,1/(2 s^2)}}
\left(\frac{x_0}{x}\right)^s \; \frac{d x}{x}\;e^{\lambda_+(s) \,\chi}
\;,\;\;\;\;\;\;
\chi(t) \;=\; -\frac{1}{\lambda_+^\prime(s)}\,
\left[\ln\left(\frac{x_0}{x}\right)\;+\;\frac{1}{2 s}\right]
\nonumber
\\
g(x, t) \, dx&=&  \frac{1}{\sqrt{2 \pi}}\,
\frac{H_-(s)}{\sqrt{\lambda_+^{\prime\prime}(s) \, \chi}}
\left(\frac{x_0}{x}\right)^s \; \frac{d x}{x}\;e^{\lambda_+(s) \,\chi}
\;,\;\;\;\;\;\;
\chi(t) \;=\; -\frac{1}{\lambda_+^\prime(s)}\,\ln\left(\frac{x_0}{x}\right)
\nonumber
\\
I_q(x, t) &=&  \frac{1}{\sqrt{2 \pi\, s}}\,
\frac{J(s)}{\sqrt{\lambda_+^{\prime\prime}(s) \, \chi\,+\,1/(2 s^2)}}
\left(\frac{x_0}{x}\right)^s \; e^{\lambda_+(s) \,\chi}\;,\;\;\;\;\;\;
\chi(t) \;=\; -\frac{1}{\lambda_+^\prime(s)}\,
\left[\ln\left(\frac{x_0}{x}\right)
\;-\;\frac{1}{2 s}\right]
\nonumber
\\
I_g(x, t) &=&  \frac{1}{\sqrt{2 \pi\, s^2}}\,
\frac{H_-(s)}{\sqrt{\lambda_+^{\prime\prime}(s) \, \chi\,+\,1/s^2}}
\left(\frac{x_0}{x}\right)^s \; e^{\lambda_+(s) \,\chi}\;,\;\;\;\;\;\;
\chi(t) \;=\; -\frac{1}{\lambda_+^\prime(s)}\,
\left[\ln\left(\frac{x_0}{x}\right)
\;-\;\frac{1}{s}\right]
\;\;.
\label{gspec}
\end{eqnarray}
Recall that the time evolution variable $\chi(t)$,
defined in (\ref{chi}), relates
these spectra to real time $t$ via eq. (\ref{chi2}),
so that $t$ can be resubstituted if desired, provided $v_\tau$ in (\ref{vtau})
is known.
I would like to remind the reader that one of the advantages for using
$\chi$ , eq. (\ref{chi}), as evolution variable rather than the laboratory
time $t$, was to absorb the nuclear medium dependence and thereby obtain
a relatively model independent description. Thus, notwithstanding the
uncertainty of medium effects, the properties of the parton shower
evolution are essentially controlled by perturbative QCD, but the
connection to real time resides in the relation (\ref{chi2}).
{}From this point of view,
the parton spectra (\ref{qspec}) and (\ref{gspec}) represent
the elementary solutions to the evolution equation (\ref{Schroe}) under the
approximations made in Sec. 3.1 and summarize
the longitudinal parton cascade evolution in nuclear matter
within the present framework.
It is evident that the variation of the number of shower particles
with time follows an exponential law with exponent $\lambda_+(s) \chi(t)$
with the (implicitely medium dependent)
'absorption coefficient' $(-\lambda_+)$ describing the variation
with time and the power $(-s)$ controlling the momentum distribution
in $x_0/x$.

It is worth noting that the distributions $\wp$ and $g$ reduce to the
well known approximate solutions to the $Q^2$-evolution
according to the DGLAP equations
\cite{lla3}, if one sets the nuclear matter density equal to zero.
Then $\tau = v_\tau t = t$ or $v_\tau =1$, because in the absence
of scattering and fusion processes the particles can only evolve
by successive branchings with the age $\tau$ being equal to laboratory time
$t$.
The connection between time- and $Q^2$-evolution, as discussed in
Sec. 2.1, is given by the average relation (\ref{tQ}), $Q^2=xP/t$.
Then, considering the evolution of the gluon distribution
alone, at small $x$ the behavior
is controlled by the branching kernel $F_{gg}(s)=C(s)$
near $s=0$ [c.f. eq. (\ref{bk})].
Hence, taking only the dominant term in $F_{gg}$
into account, one has
\begin{equation}
\frac{\partial}{\partial t} \,G \;=\; -\,F_{gg}\;G
\;\;,\;\;\;\;\;\;\;\; F_{gg}(s)\;\approx\; -\frac{2 C_G}{s} \;=\;-\lambda_+(s)
\;\;,
\end{equation}
where $C_G = n_c$ as before.
Then the saddle point evaluation yields (setting $x_0=1$):
\begin{equation}
g(x,t)\,dx\;=\;\frac{1}{2\pi}\;
\frac{\left[\pi \,\frac{n_c}{b}\,
\ln\left(\frac{\alpha_s(Q^2)}{\alpha_s(Q_0^2)}\right)
\,\ln(1/x)\right]^{1/4}}{\ln(1/x)}
\,\exp\left[\sqrt{\frac{4}{\pi} \,\frac{n_c}{b}
\,\ln\left(\frac{\alpha_s(Q^2)}{\alpha_s(Q_0^2)}\right)
\,\ln(1/x)}\right]
\;\frac{dx}{x}
\;\;,
\label{gg0}
\end{equation}
where the time variable $\chi$, when re-expressed in terms of $Q^2$ rather
than $t$, is given by
\begin{equation}
\chi\left(Q^2\right)\;=\;
\frac{1}{2 \pi b}\;\ln\left(\frac{\alpha_s(Q^2)}{\alpha_s(Q_0^2)}\right)
\;=\;\frac{1}{2 \pi b}\;
\ln\left(\frac{\ln(Q_0^2/\Lambda^2)}{\ln(Q^2/\Lambda^2)}\right)\;\;,
\label{chi0}
\end{equation}
with $Q_0^2=P/t_0$ and $Q^2=xP/t$ ($t>t_0$).
Eq. (\ref{gg0}) together with (\ref{chi0}) coincides with leading logarithmic
behavior of the gluon distribution within the LLA \cite{lla3}.
\smallskip

Let me return now to the parton spectra (\ref{qspec}) and (\ref{gspec}), and
examplarily exhibit some characteristic features of the
longitudinal parton shower evolution in nuclear matter, as reflected by the
$\chi$, $x$ and nuclear ($A$) dependence of the spectra.
In what follows I
chose the virtuality of the cascade initiating parton $Q_0 = 30$ GeV.
An increase (decrease) of $Q_0$ generally results in a larger (smaller) overall
multiplicity of secondaries and also shifts the evolution toward
earlier (later) times.
The initial longitudinal momentum fraction was set to $x_0 = 1$, since the
shower development depends only on the ratio $x_0/x$.
Finally the QCD scale $\Lambda$ that enters the running coupling
$\alpha_s$ was taken equal to 230 MeV.
In order to exhibit the nuclear medium influence
on the shower development, as controlled by eq. (\ref{htau}),
I considered as before the three cases: a) $A=0$ (absence of
medium), b) $A=50$ (moderately dense medium), and, c) $A=200$ (dense medium).
\smallskip

In Figs. 4 and 5 the parton number densities $\wp(x,t)$ and $g(x,t)$
from (\ref{qspec}) and (\ref{gspec}), respectively, are plotted
for $x=10^{-1}$ and $x=10^{-3}$ versus $\chi$ (it is helpful
to keep in mind that for slowly varying $\alpha_s$, one has
approximately $\chi \propto \ln t$).
The essential observations are
basically the same for both  quark and
gluon initiated showers:

(i) For a given value of the ratio $x$, the change with "time"
$\chi$ of the spectra $\wp$ and $g$ reflects the aging of the shower.
The multiplicity initially increases
({\it young shower}), goes through a peak ({\it shower maximum}),
and then decreases again ({\it old shower}).
The time it takes for a parton shower to evolve from $x_0=1$ downwards
to some value $x$ naturally increases with decreasing $x$ and at the same
time the average multiplicity of produced particles grows rapidly.
(The integral spectra $I_q$ and $I_g$, which are not shown here, exhibit
an  analogous behaviour.)

(ii) Comparing the cases a) to c), the influence of the nuclear
medium manifests itself in a slowing down of the shower development,
particularly at later times
($\chi \,\lower3pt\hbox{$\buildrel > \over\sim$}\, 2$).
In other words, the aging of the cascading partons is more
delayed, the denser the medium is. Related to that is the
clear increase of the parton multiplicity
at a given $x$ with larger nuclear density, because
repeated energy-momentum transfers by scatterings and fusions
rejuvenate the shower partons and amplify the bremsstrahlung processes.
\smallskip

In order to exhibit the $x$-dependence of the parton spectra at fixed
value of $\chi$ let me define the so-called {\it normal spectra} \cite{cr1} at
the
shower maximum. The normal spectra correspond to an {\it optimum value}
$\chi = \chi_0$, which coincides with the
value of $\chi$ that makes the function $\exp[\phi(s)]$ of eq. (\ref{phi})
a maximum, because the other terms change slowly with $\chi$.
Thus, $\chi_0$ is defined by the equation
\begin{equation}
\frac{\partial}{\partial \chi}\,\exp[\phi(s)]\;=\;0
\;\;\;\;\;\;\; \mbox{at $\chi=\chi_0$}
\end{equation}
or
\begin{equation}
\lambda_+(s)\;+\;\left(y \,+\,\lambda_+^\prime(s) \chi \,-\,\frac{k}{s}\right)
\;\left(\frac{\partial s}{\partial t}\right)_{\chi=\chi_0}\;=\;0
\;\;.
\label{optchi}
\end{equation}
Recalling that eq. (\ref{saddle}) is satisfied, one has
\begin{equation}
\lambda_+(s)\;=\;0
\;\;\;\;\;\;\;\mbox{at}\; \chi\;=\;\chi_0
\;\;.
\end{equation}
Hence, at $s = s_0$,
\begin{equation}
\chi_0\;=\; -\,\frac{y\;-\;k}{\lambda_+^\prime(s_0)}
\;=\;
 -\,\frac{1}{\lambda_+^\prime(s_0)}\;
\left[ \ln\left(\frac{x_0}{x}\right) \;-\; k \right]
\;\;.
\end{equation}
The normal spectra of $\wp$, $g$, $I_q$, $I_g$ and the
corresponding optimum values $\chi_0$ are easily obtained
from eqs. (\ref{qspec}) and (\ref{gspec}) by putting $s=s_0$, $\lambda_+=0$,
and $\chi=\chi_0$. The values of $s_0$, given by
$\lambda_+(s_0)=0$, and
of $\lambda_+^\prime(s_0)$ can be read off Fig. 2.

The top part of Fig. 6 displays the typical shape of the normal spectrum
around shower maximum, corresponding to $s=s_0$ and $\chi = \chi_0$.
It is interesting that, when plotted as a normalized probability distribution
$p(\chi_0)$, this normal spectrum is unique: first, it is the
same for quarks and gluons, second, it is independent of the type
of primary particle, and, third, it
does not depend on the nuclear medium density.
On the other hand, the corresponding optimum values $\chi_0$ vary
significantly for the different cases, as
is evident from the lower part of Fig. 6, which shows $\chi_0$ versus $x$ for
the three aforementioned cases a) $A=0$, b) $A=50$ and c) $A=200$.
This behaviour is to be expected,
since $\chi_0$ marks the characteristic time that it takes for the parton
shower
to build up the normal spectrum.
Looking at Fig. 6, one can conclude:

(i) With decreasing $x$ the values of $\chi_0$ increase, roughly as $\ln(1/x)$.
The slope is the same for both quarks and gluons, but $\chi_0^g$ is always
smaller than $\chi_0^q$, reflecting the fact that the gluons
evolve faster due to their larger interaction and radiation probability.

(ii) The effect of the nuclear medium is again a substantial dilation of
the shower development, and thus  toward the establishment of the normal
spectrum. This is clearly reflected by the steepening of the slope
of the time scale $\chi_0$ when proceeding from the free space case a) to
the dense matter case c).
\bigskip

\noindent {\bf 5. SUMMARY}
\medskip

In this paper I presented an analytical method to solve the QCD evolution
equations derived in Ref. \cite{gm1}
which describe the cascading of quarks and gluons in a nuclear
environment. These evolution equations extend the well known DGLAP
equations for jet development in hard QCD processes, to the time
evolution of parton cascades in a nuclear medium, including in addition
to spontaneous branchings also fusion, scattering and stimulated branching
processes.
In order to obtain tractable closed solutions to the evolution equations,
a number of idealizing approximations were necessary.
Therefore, at this point, this approach does of course not provide quantitative
predictions of parton transport phenomena in a dense matter environment.
The main uncertainty is here the lack of detailed knowledge
of the form of the medium influence on the cascade evolution.
Since frequency and type of interactions of the partons with the medium must
depend crucially on the nuclear density and structure, one must supply some
external input. In order to minimize the amount of phenomenological input,
the medium influence was therefore hidden into a time evolution
variable $\chi(t)$, the relation of which with the laboratory time $t$
however is determined by the specific properties of the nuclear medium.
As advantagous trade-off one gains that
the calculated properties and spectra of the
cascading partons are a rather firm consequence of perturbative
QCD proceses in a nuclear medium being
characterized by a specific time variable $\chi$.

Of prime interest in this work was the multiplicative behaviour
of cascading partons at small $x$, since such soft particles each take away
only a negligable portion of the total energy, but form at the same
time the bulk of multiplicity.
The characteristic properties of parton shower evolution inside
nuclear matter with respect to the variables time ($\chi$), longitudinal
momentum fraction ($x$) and nuclear density ($A$) can be summarized as follows:
\begin{description}
\item[(1)]
With respect to the rate of time change at fixed value of $x$,
a typical parton shower can be divided roughly  in
three stages:
(i)
{\it young shower}:
during the early stage the shower particles rapidly multiply
due to branching processes (mainly gluon emission);
(ii)
{\it shower maximum}:
for a short time the number of particles
in the cascade remains almost constant;
(iii)
{\it old shower}:
the increasing dissipation of energy due to branchings
as well as a growing number of fusion and scattering processes
reduce the number of shower particles at given $x$ during
later times.
\item[(2)]
The mean of the $x$-distribution of shower partons shifts with progressing time
to smaller values of $x$, but this development is significantly delayed
in a dense matter environment as compared to free space. The
$x$-spectra exhibit a unique form around the shower maximum, independent
of type of parton and of the nuclear density. This {\it normal spectrum}
is established after a certain {\it optimum time} $\chi=\chi_0$, which measures
the characteristic time scale that depends on the
density of the surrounding medium: the denser the matter,
the longer it takes for the partons to build up to their normal spectrum.
Also, gluons always reach this point earlier than quarks, because of their
larger cross-section and radiation probability.
\end{description}

As said before, the approach presented here has no
truly predictive relevance for experimental observables yet.
Rather than that it may be a first step
towards a more fundamental understanding of "QCD in medium",
using well developed methods of renormalization group
improved perturbation theory for "QCD in vacuum".
I believe that the most important issues to be addressed in the near future
within such a program are the following:
\begin{description}
\item[(1)]
The rigorous distinction between the shower particles and
the nuclear partons needs to be dropped. Instead both should
be treated on the same footing in order to describe a self-consistent dynamical
multi-parton system. This would then incorporate also time- and $x$-dependent
nuclear parton distributions that receive modifications due to
the response of the shower evolution.
An interesting aspect of such a dynamical interplay between scatterings,
multiplication and recombination of partons is the question
of a possible saturation \cite{saturation} of the
parton number densities that may render
a finite parton density as $x \rightarrow 0$.
\item[(2)]
In the context of nuclear structure, the relevance and the form of
the age distribution ${\cal A}$ of vitual shower partons needs to be studied,
because it provides an interface between
the nuclear medium and the response of the cascading partons to this
environment.
Therefore ${\cal A}$ may be experimentally accessible by comparing
inclusive quantities associated with parton evolution in e.g. deep
inelastic scattering experiments on hadrons versus heavy nuclei.
This issue is of great interest from both theoretical as well as
experimental side.
\item[(3)]
Interference effects which are important particularly in
the small $x$-domain need to be taken into account systematically. These
include two well known phenomena, namely,
a) the coherence of successive soft gluon emissions in the
parton cascade leading to destructive interference and
a suppression of gluon radiation, and,
b) the Landau-Pomerantchuk-Migdal effect, i.e. the
interference of induced gluon emission amplitudes in multiple parton
collisions,
which also causes a depletion  in the gluon radiation spectrum.
\item[(4)]
A possible extension to include the time development of the transverse
momentum distribution of partons as well as the lateral spread of
shower particles would be desirable. This would then provide
a solution to the cascade development in full four dimensional momentum
space, in virtuality $Q^2$, longitudinal and transverse momentum
components $x$ and $k_\perp^2$, respectively.
\item[(5)]
Parallel to the previous items a comparison with a full
six-dimensional phase-space analysis, as e.g. possible with
Monte Carlo simulations \cite{pcm0}, is of great importance to test the
accuracy of the analytical solutions. Such a twofold investigation
can be  explorative for both analytical and numerical solutions
of the evolution equations, in the sense that mutual feedback
may lead to an improved understanding of the microscopic
parton dynamics in nuclear matter.
\end{description}
\smallskip

\bigskip

\section*{Acknowledgements}
\vspace{-2mm}
\noindent
I would like to thank H.-T. Elze for many inspiring conversations
during the course of this work.
\bigskip

%\newpage

\newpage

{\bf FIGURE CAPTIONS}
\bigskip

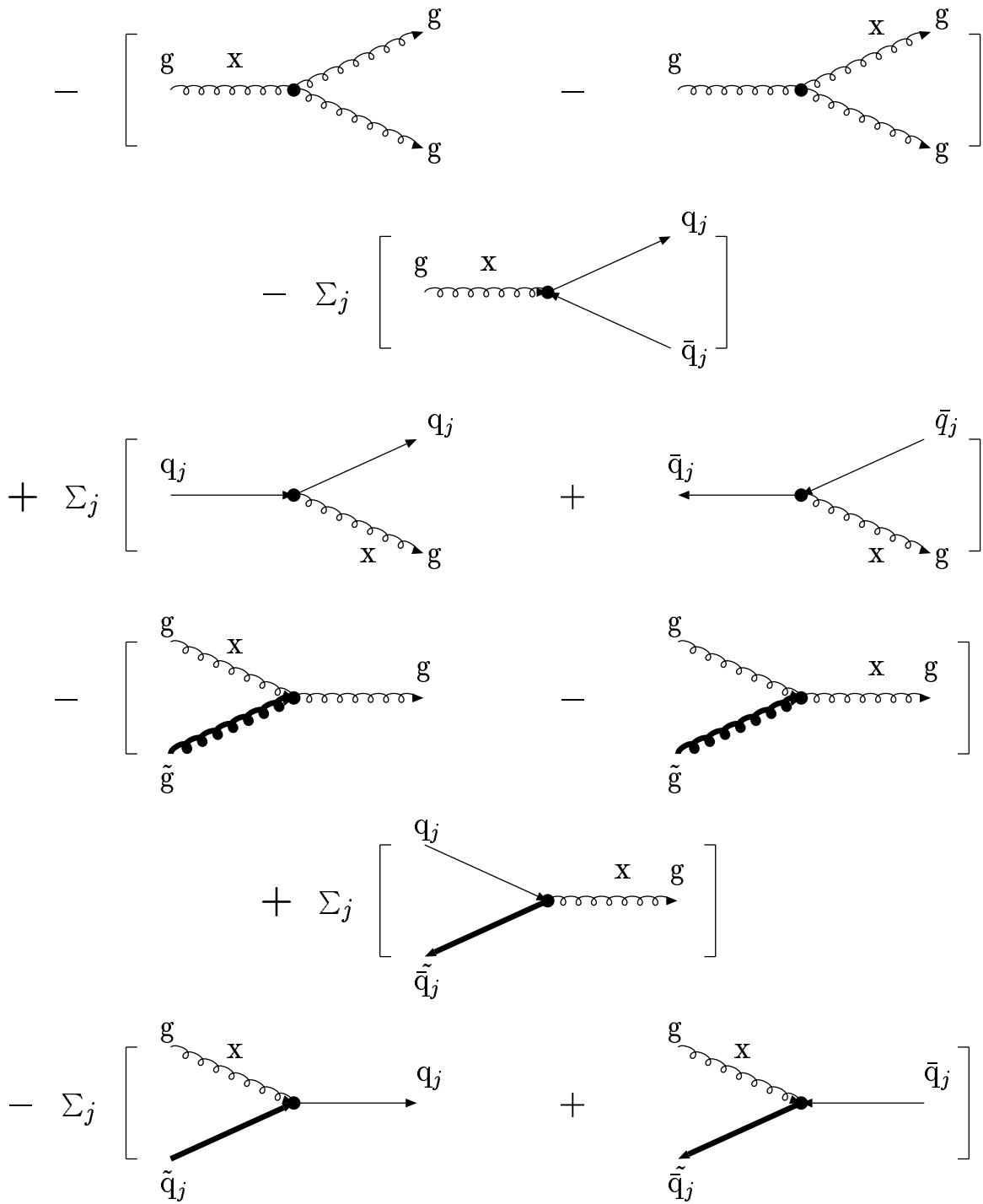
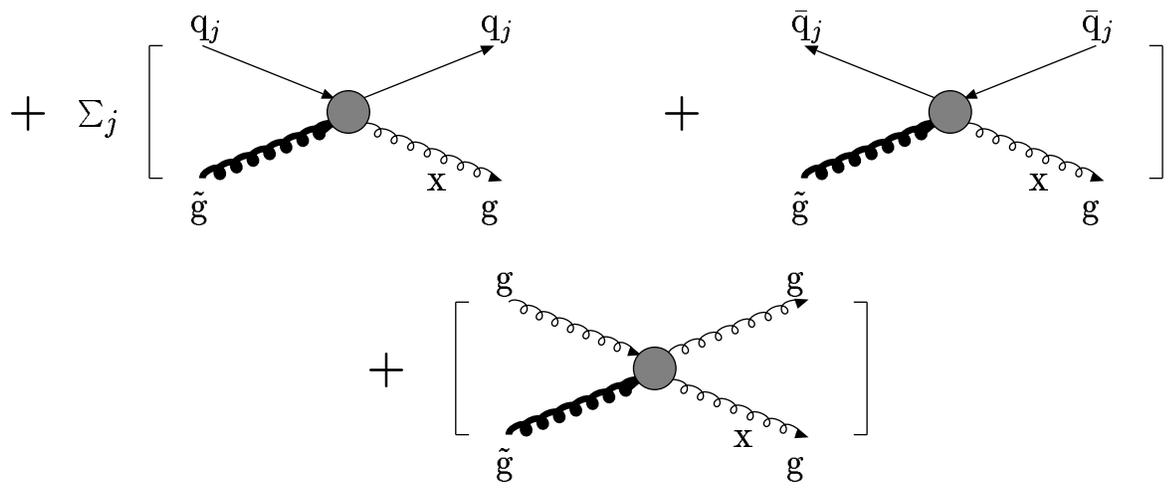
\noindent {\bf Figure 1:}
Diagrammatic representation of the coupled evolution equations
(\ref{eq})-(\ref{eg}) for a) quarks (and similarly antiquarks) and b) gluons.
In contrast to the branching processes, the fusion and scattering processes
involve interactions of a cascade parton with a nuclear parton (marked by
thick lines).
In a) the various diagrams correspond to the integral operators
$-\hat A Q_i$, $\hat B G$, $-\hat A^\prime Q_i$, $-\hat B^\prime Q_i$,
$\sum_j [\hat S_{qq} Q_j + \hat S_{q\bar q} \bar Q_j]$,
$\hat S_{qg} G$, and in b) to
$-\hat C G$, $-\hat D G$, $\sum_j [\hat E Q_j + \hat E \bar Q_j]$,
$-\hat C^\prime G$, $-\hat D^\prime G$,
$\sum_j [\hat E^\prime Q_j + \hat E^\prime \bar Q_j]$,
$\sum_j [\hat S_{gq} Q_j+ \hat S_{g\bar q} \bar Q_j]$,
$\hat S_{gg} G$.
\medskip

\noindent {\bf Figure 2:}
Form of the functions
$F_{qq}, F_{qg}, F_{gq}, F_{gg}$, eq. (\ref{fab}),
with respect to the variable $s$.
The three plots a)-c) correspond to different nuclear density of the medium
measured in terms of nuclear mass number: $A=0$
(free space), b) $A=50$ (moderately dense medium),
and c) $A=200$ (dense medium).
\medskip

\noindent {\bf Figure 3:}
Functional behavior of
$\lambda_+, \lambda_-$, eq. (\ref{lambda+-}),
and of the derivatives $\lambda_+^\prime$, $\lambda_+^{\prime\prime}$
versus the variable $s$.
The cases a)-c) correspond to Fig. 2.
\medskip

\noindent {\bf Figure 4:}
{\it Primary quark}:
Time-evolution of the differential spectra for the parton number densities
$\wp(x,t)$ and $g(x,t)$ with respect to the variable $\chi(t)$,
according to (\ref{qspec}), at fixed values
$x=10^{-1}$ and $x=10^{-3}$.
The virtuality of the cascade initiating quark was chosen $Q_0=30$ GeV.
Cases a)-c) as in Fig. 2.
\medskip

\noindent {\bf Figure 5:}
{\it Primary gluon}:
Development with time $\chi(t)$ of the parton number densities
$\wp(x,t)$ and $g(x,t)$ for the case of a shower initiating gluon,
eq. (\ref{gspec}),  with initial virtuality
$Q_0=30$ GeV and at fixed values
$x=10^{-1}$ and $x=10^{-3}$.
Cases a)-c) as in Fig. 2.
\medskip

\noindent {\bf Figure 6:}
{\sl Top}:
Unique shape of the {\it normal spectrum}
around shower maximum in terms of the normalized probability distribution
$p(\chi_0) = a(x,\chi_0)/\int dx a(x, \chi_0)$ where $a=\wp, g$.
{\sl Middle and bottom}:
The {\it optimum values} $\chi_0$ for gluons and quarks
versus $x$, corresponding to the
point of time $\chi(t)$ when the normal spectrum at a certain
value of $x$ is established. The three sets of curves a) and c) refer to
different nuclear density, as in Fig. 2. The virtuality of the primary
quark, respectively gluon, was chosen $Q_0=30$ GeV.
\medskip

\vfill

\begin{thebibliography}{99}
\bibitem{comment1}
There is recent experimental evidence from HERA data that
at very small $x \,\lower3pt\hbox{$\buildrel < \over\sim$}\, 10^{-3}$
the gluon structure function of a proton exhibits a "Lipatov enhancement"
$x \hat g(x) \propto x^{-C}$ with $c\approx 1/2$. Nevertheless, for
reasons of simplicity mentioned in the text, I will use the approximation
of a constant behaviour at small $x$, which corresponds to $C=0$.
\end{thebibliography}
\end{document}